\documentclass[journal]{IEEEtran}

\usepackage{graphicx}
\usepackage{cite}
\usepackage{picinpar}
\usepackage{amsmath}
\usepackage{url}
\usepackage{flushend}
\usepackage[latin1]{inputenc}
\usepackage{colortbl}
\usepackage{soul}
\usepackage{multirow}
\usepackage{pifont}
\usepackage{color}
\usepackage{alltt}
\usepackage[hidelinks]{hyperref}
\usepackage{enumerate}
\usepackage{siunitx}
\usepackage{breakurl}
\usepackage{epstopdf}
\usepackage{pbox}

\usepackage{amssymb,amsmath,verbatim,bm,color,times,mathtools,multicol,relsize}
\usepackage{float}
\usepackage{subcaption}
\usepackage{siunitx}

\newcommand{\matrixstyle}[1]{\mathrm{#1}}
\newcommand{\vectorstyle}[1]{\boldsymbol{\mathrm{#1}}}

\begin{document}

\title{ Neural Network Augmented Physics Models for Systems with Partially Unknown Dynamics: Application to Slider-Crank Mechanism}

\author{Wannes De Groote, Edward Kikken, Erik Hostens, Sofie Van Hoecke, Guillaume Crevecoeur %
\thanks{W. De Groote holds a doctoral grant strategic basis research  (3S07219) of the Fund for Scientific research Flanders (FWO). This work was furthermore supported by Flanders Make, the strategic research centre for the manufacturing industry, through the Flanders Make ICON MODA project. Additionally, this research received funding from the Flemish Government under the "Onderzoeksprogramma Artificiële Intelligentie (AI) Vlaanderen" program.}
\thanks{W. De Groote and G. Crevecoeur  are with the Department of Electromechanical, Systems and Metal Engineering, Ghent University, 9000 Ghent, Belgium, and also with EEDT-DC, Flanders Make, 3920 Lommel, Belgium (e-mail: wannes.degroote@ugent.be; guillaume.crevecoeur@ugent.be).}
\thanks{S. Van Hoecke is with the Internet Technology and Data Science Lab (IDLab) of Ghent University and imec, 9000 Ghent, Belgium (e-mail: sofie.vanhoecke@ugent.be).}%
\thanks{E. Kikken and E. Hostens are with the Core Lab DecisionS, Flanders Make, 3920 Lommel, Belgium (e-mail: edward.kikken@flandersmake.be; erik.hostens@flandersmake.be).}%
}

\maketitle

\begin{abstract}
Dynamic models of mechatronic systems are abundantly used in the context of motion control and design of complex servo applications. In practice, these systems are often plagued by unknown interactions, which make the physics-based relations of the system dynamics only partially known. This paper presents a neural network augmented physics (NNAP) model as a combination of physics-inspired and neural layers. The neural layers are inserted in the model to compensate for the unmodeled interactions, without requiring direct measurements of these unknown phenomena. In contrast to traditional approaches, both the neural network and physical parameters are simultaneously optimized, solely by using state and control input measurements. The methodology is applied on experimental data of a slider-crank setup for which the state dependent load interactions are unknown. The NNAP model proves to be a stable and accurate modeling formalism for dynamic systems that ab initio can only be partially described by physical laws. Moreover, the results show that a recurrent implementation of the NNAP model enables improved robustness and accuracy of the system state predictions, compared to its feedforward counterpart. Besides capturing the system dynamics, the NNAP model provides a means to gain new insights by extracting the neural network from the converged NNAP model. In this way, we discovered accurate representations of the unknown spring force interaction and friction phenomena acting on the slider mechanism.

\end{abstract}


\IEEEpeerreviewmaketitle

\section{Introduction}

\IEEEPARstart{S}{ervo}-controlled systems face increasingly demanding performance and efficiency requirements in industrial and manufacturing applications. In, for example, presses, pumps and compressors, linear motion must be realized in direct drive or indirectly. The latter servo systems need to drive a load system via a rotary motor system and mechanical transmissions such as gears, cam-follower, and bar linkages. Position and speed control are critical, and their performance is limited due to unknown dynamics resulting from the external loads and uncertainties in the mechanical system such as inertia, backlash, damping, and friction \cite{kim2018moment, papageorgiou2018robust, wang2019adaptive}. To better understand these intricate and often highly nonlinear dynamics, extensive research has been conducted on developing predictive models that capture the overall behavior of servo systems. These models can then be used by a user or manufacturer to further improve the mechatronic system design \cite{kamadan2017co, kim2015feasibility}. Furthermore, predictive models can be inserted in model-based control strategies to improve the mechatronic system's operations, (e.g. \cite{ziyuan2019An}). 

Capturing the servo drive system dynamics is, however, often cumbersome and challenging because we face against limits of various existing modeling and system identification formalisms. The modeling of a mechatronic system can be based on expert knowledge where partial and/or ordinary differential equations are distilled from the system. These physics-based models include (lumped) physical parameters that relate directly to the actual mechatronic system's behavior. Parameter identification techniques can then be used to identify the values of these parameters based on experimental sensor data \cite{schon2011system}. Uncertainties can, however, still remain because servo drive systems face nonlinearities such as friction forces that give rise to disturbances that are difficult to model. Alternatively, black-box models can be built in a data-driven manner. In the context of nonlinear system identification, several methodologies exist such as  Gaussian processes \cite{ kocijan2016modelling}, Hammerstein-Wiener structures \cite {wills2013identification}, and nonlinear auto-regressive models (e.g. NARMAX) \cite{ billings2013nonlinear}. Supervised learning is typically used to build the model by relating input to output data. Supervised machine learning techniques such as deep learning have become pervasive because they have the ability to recognize patterns in data \cite{lecun2015deep}. 
Neural networks have, for instance, been used to approximate various nonlinear relations, such as inverse kinematics within complex mechatronic systems \cite{lakhal2015hybrid}. Recurrent neural networks (RNN) have the ability to capture the nonlinear dynamics in time series by means of an internal state and are of particular interest for learning the dynamic behavior of mechatronic applications\cite{ogunmolu2016nonlinear}. Ensembles of interconnected RNNs have demonstrated their usefulness in the modeling of complex dynamical systems \cite{mohajerin2014modular}. However, as observed in \cite {mohajerin2019multistep}, there is no physical interpretation given to this internal state, which makes the convergence rely substantially on hyperparameter tuning and initialization choices.

Having physical interpretations to the modeling can, however, be useful, particularly to assure robust and physics consistent predictions. Physically interpretable neural-fuzzy networks have shown in this respect enhanced explainability by assigning piecewise linear models to the network nodes  \cite{liu2019identification}. The dynamic behavior is typically defined by the so-called derivative function in a partial and/or ordinary differential equations-based model. Based on physical insights, it is possible to deduce an appropriate structure of a black-box model. Feedforward neural networks have been proposed in physics-informed deep learning \cite{raissi2018multistep} to approximate derivative functions of dynamical systems that, in contrast to traditional black-box approximations, return more interpretable models. Other recent approaches based on symbolic regression \cite{schmidt2009distilling} and sparse regression \cite{brunton2016discovering} were able to return more interpretable models by discovering the underlying governing equations. Furthermore, recent developments indicate the possibilities to approximate reformulations of classical mechanics, such as Lagrangian \cite{lutter2019deep} and Hamiltonian \cite{greydanus2019hamiltonian} mechanics by neural networks. 

Next to using physical insights to deduce the structure of black-box models, one can directly use physics-based governing equations within the data-driven model. Physical laws can be included in the loss function of black-box neural networks to guide the training process towards physical consistent model predictions \cite{jia2019physics}. Conversely, neural network mappings can assist in compensating for prediction discrepancies of a complete physics-based model \cite{mohajerin2019multistep,ajay2019combining}. These neural networks serve as nonlinear mappings that transform the state estimations behavior of the physics-based model. They enable predictions to be improved; however, they do not represent an interpretable physical phenomenon. An incomplete physics model was considered in \cite{punjani2015deep} that was complemented with neural networks. The  latter was engaged to accommodate for partially unknown dynamics within the physical system. Neural networks were trained separately and in a supervised manner on the basis of measurement data that directly relate to the unknown phenomena. 

\begin{figure}[H]
	\centering
	\vspace{-10pt}
	\includegraphics[trim=5.6cm 6.3cm 9cm  12cm,clip=true, angle =-90, width=0.7\columnwidth ]{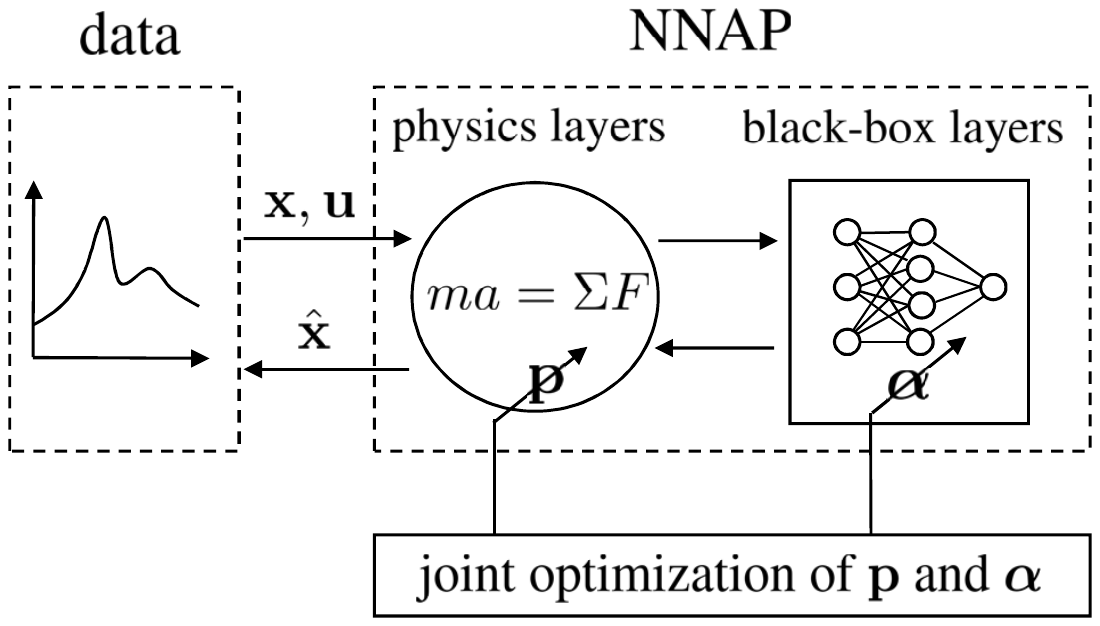}
	\caption{NNAP model as an interconnected network of physics and black-box layers, which are defined by trainable physics parameters $\vectorstyle{p}$ and neural parameters $\vectorstyle{\alpha}$, respectively. All layers are simultaneously optimized solely by using control inputs $\vectorstyle{u}$ and state measurement $\vectorstyle{x}$.}
	\vspace{-10pt}
	\label{fig:jointOptim}
\end{figure}

This paper proposes a neural network augmented physics (NNAP) model, encompassing trainable physical layers closely connected to a neural network to compensate for partially unknown dynamics. It is assumed that no direct measurement data on the unknown phenomena are available because they are either difficult or not possible at all to measure. Examples are unknown friction phenomena in servo mechanisms that affect the overall system behavior. The proposed NNAP models consist of an interconnected network of physics layers complemented with black-box layers. Measurement data are propagated through these layers to simultaneously identify the physics and data-driven parameters, as depicted in Fig. \ref{fig:jointOptim}. The effectiveness of the methodology is extensively tested on experimental data collected on a servo slider-crank setup with unknown nonlinear load and unidentified key physical parameters. The objective of the presented methodology is twofold. First, to obtain a model with improved predictive capabilities by performing a joint optimization of physical and neural parameters while only using state and control input data. Secondly, to explain the partially unknown dynamics within the system by analyzing the information captured by the black box layers after training the NNAP model. By using such an approach, new physical insights can be obtained to explain and interpret the unknown phenomena that are inaccessible by direct measurements or are difficult to model \textit{ab initio}.

\section{Slider-Crank System} \label{sec:SCsystem}
Various industrial applications require reciprocating linear motions. Motion control of these systems can be provided by means of linear electric motors \cite{boldea2017linear}. For high-power applications, a combination of rotary motors in combination with mechanical transmissions \cite{sharma2006theory} such as gears, cam-follower and bar linkage systems are the preferred driveline because they achieve energy efficient and robust operation. These drivelines exhibit highly nonlinear behavior that is often challenging to capture using first-principle physics modeling. They are often plagued by unidentified load disturbances and unknown interactions of the system with the environment that are beyond expert knowledge. These nonlinear phenomena are, henceforth, formalized by relation $\mathcal{P}$ that relates certain inputs of the system (e.g. velocity) to a response of the system (e.g. friction force). Next to these nonlinear unknown phenomena, physical parameters such as the inertia in the servo drive system can be uncertain and are formalized as $\vectorstyle{p}$.\\
This paper applies and aims at validating a methodology to accommodate for unknown $\mathcal{P}$ and $\vectorstyle{p}$ that can arise in servo-controlled systems. We consider a slider-crank mechanism translating a rotary motion into a linear displacement that is subject to an unidentified load. This application is of direct relevance in various industrial applications such as compressors \cite{gao2010filter}, hydraulic pumps \cite{li2019design}, weaving looms \cite{eren2005comparison}, and presses \cite{zheng2014modeling}. A system model comprehending the dynamics of the slider-crank system is built, and experimental data are collected from a practical setup.  
\begin{figure}[H]
	\vspace{-5pt}
	\centering
	\includegraphics[trim=6.1cm 8.7cm 8cm  8.2cm,angle=-90,clip=true,width=0.6\columnwidth]{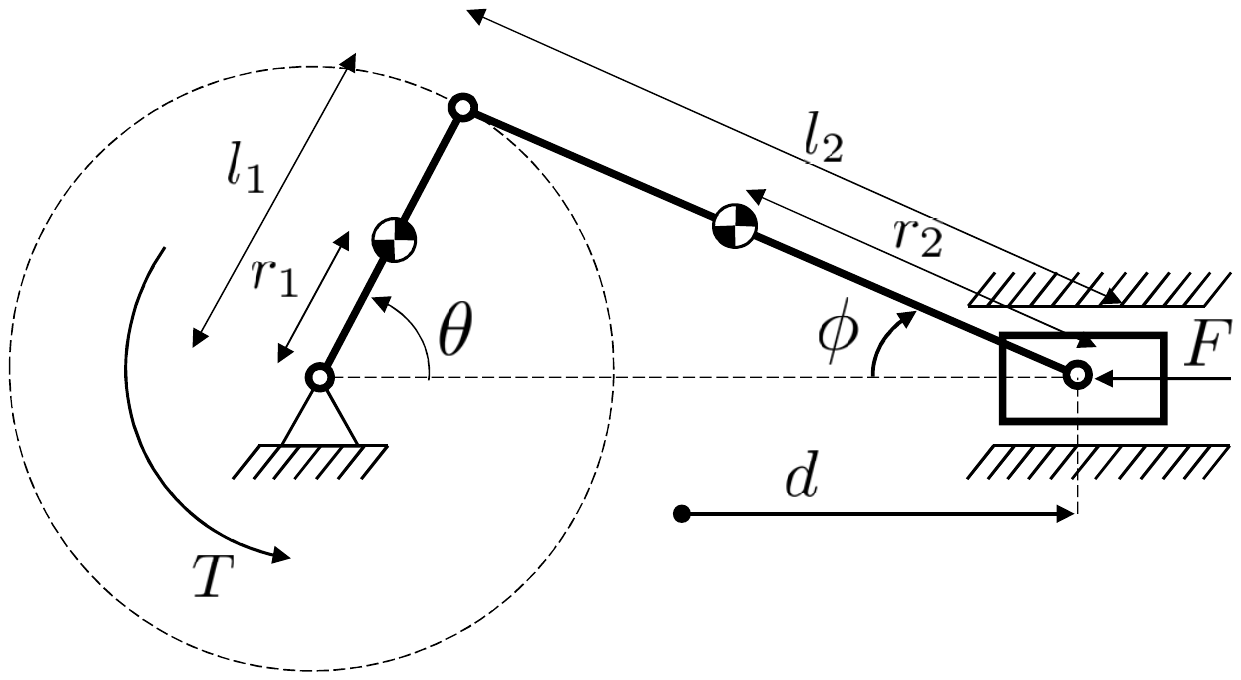}
	\caption{Schematic of slider-crank system.}
	\vspace{-5pt}
	\label{fig:geometrical}
\end{figure}
\subsection{System Model}
The slider-crank system considered in this paper is schematically depicted in Fig. \ref{fig:geometrical}. The setup consists of a rotary servo motor delivering torque $T$ to a first mechanical link that rotates with angle $\theta$, which, in turn, is connected to a second link. The latter link is connected to a slider (with relative angle $\phi$ between this link and slider). 
The dynamics of the considered slider-crank setup are governed by force interactions between the rigid mechanical links. Appendix \ref{App:Model} details the expressions that formalize the dynamics of the multibody system. A time invariant state-space model can be distilled that captures the dynamics of the state $\vectorstyle{x}=\begin{bmatrix} \theta & \omega \end{bmatrix}^T$. The angular speed is denoted as $\omega\equiv\dot{\theta}$. Note that the slider position $d$ is geometrically coupled to motor angle $\theta$ and angle $\phi$ that, in turn, is related to $\theta$ via $\phi = \arcsin\left(\frac{l_1}{l_2}\sin(\theta)\right)$. The nonlinear state space relation can be comprehended by a derivative function that relates the state $\vectorstyle{x}$ and control input $u=T$ to the next state. We, furthermore, parameterize the state space model with physical parameters $\vectorstyle{p}$, being masses, lengths, inertia, and rotational damping coefficient, which appear in (\ref{eq:Phys1}). Finally, the load force $F$ depicted in Fig. \ref{eq:geometrical} is the unknown nonlinear relation $\mathcal{P}$ that is an additional disturbance input of the model. Given the above, we formalize the dynamics in the slider-crank, thus, as $\dot{\vectorstyle{x}}=\vectorstyle{f}\left( \vectorstyle{x},u,F ; \vectorstyle{p}\right)$ with a geometrical relation between the linear displacement and velocity
\begin{equation}
\label{eq:geometrical}
\begin{aligned}
\vectorstyle{\gamma}(\vectorstyle{x};\vectorstyle{p})=\begin{bmatrix}
d \\ v
\end{bmatrix}&=\begin{bmatrix}
l_1\cos(\theta) + l_2\cos(\phi) - \left(l_2-l_1\right)\\                
-l_1\sin(\theta)\omega -  l_2\sin(\phi)\dot{\phi}  \\   
\end{bmatrix} \\          
\end{aligned}
\end{equation}
that is again dependent on the physical parameters $\vectorstyle{p}$.

\subsection{Practical Setup}
\begin{figure}[H]
	\centering
	\includegraphics[trim=3.4cm 0cm 3.7cm  0cm,clip=true, angle=-90,width=\columnwidth]{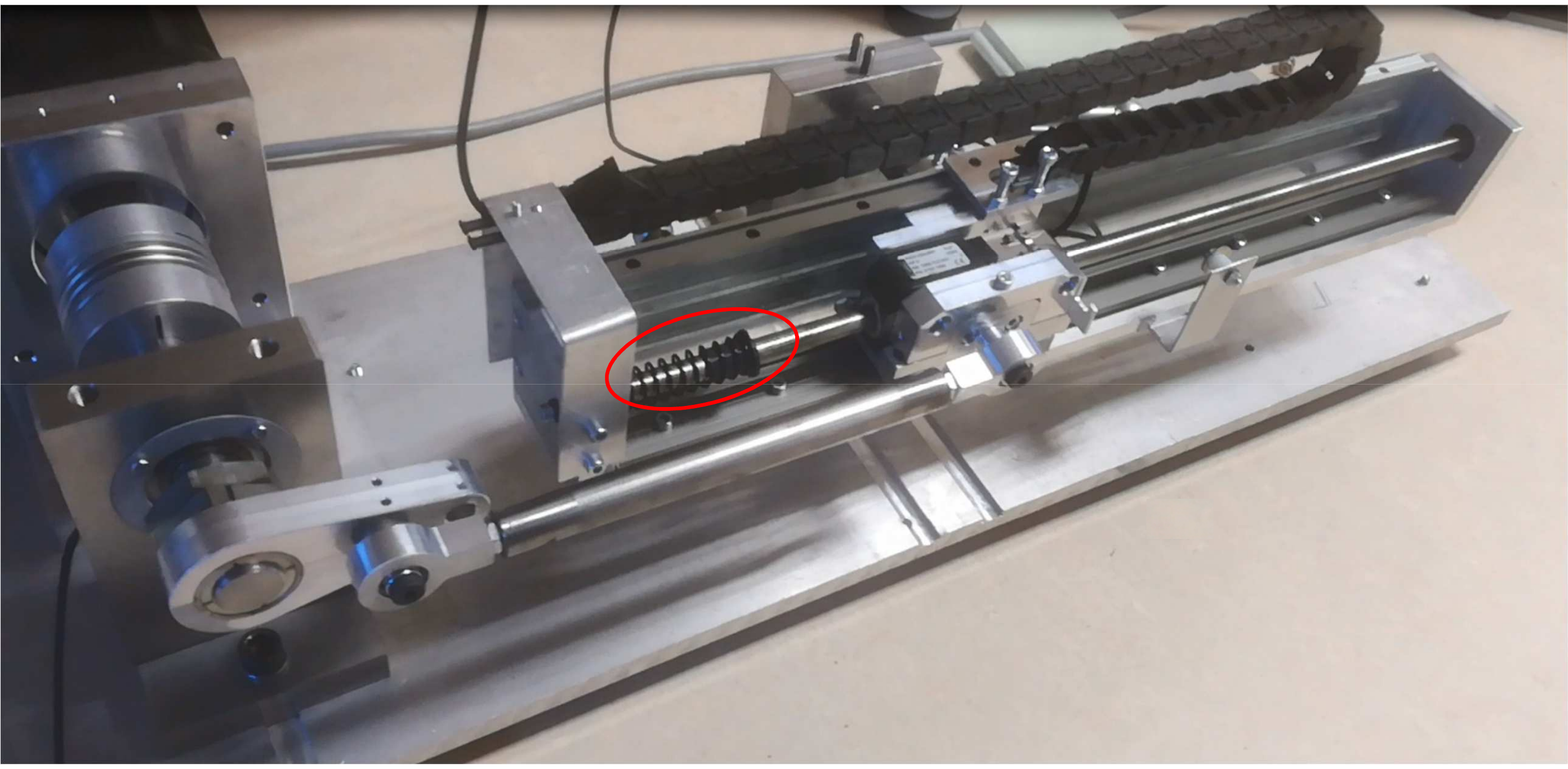}
	\caption{Slider-crank setup with spring load indicated by red circle.}
	\label{fig:setup}
\end{figure}

\begin{figure}[h]
	\vspace{0pt}
	\begin{minipage}[b]{0.45\linewidth}
		\centering
		\includegraphics[trim=6.5cm 11.9cm 7cm  12.cm,clip=true,width=1\columnwidth]{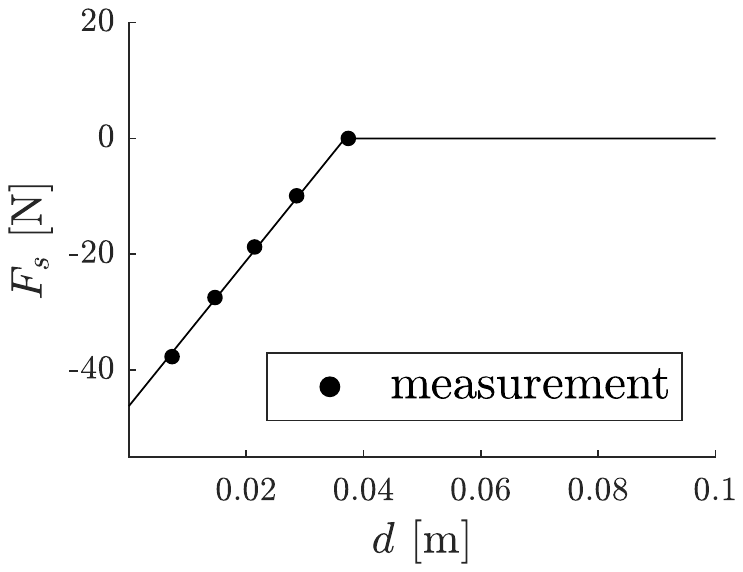}
		\vspace{-20pt}
		\caption{Spring characteristic.}
		\label{fig:springLaw}
	\end{minipage}
	\hspace{0cm}
	\begin{minipage}[b]{0.55\linewidth}
		\centering
		\includegraphics[trim=5.5cm 11.9cm 6.2cm 11.8cm,clip=true,width=1\columnwidth]{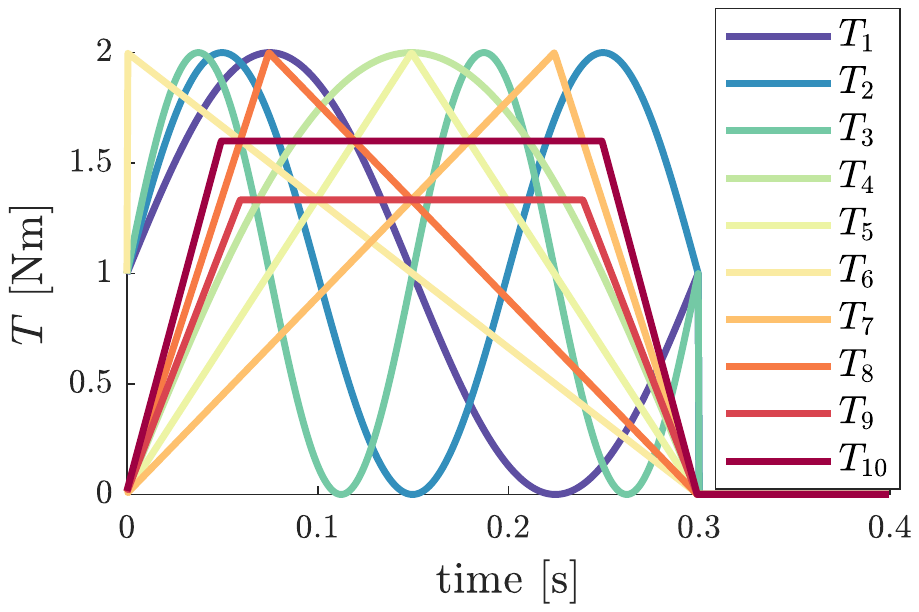}
		\vspace{-20pt}
		\caption{Torque signals.}
		\label{fig:torque}
	\end{minipage}
\end{figure}

The practical slider-crank setup is depicted in Fig. \ref{fig:setup}. A 3 \si{kW} brushless servo motor with integrated drive is connected to a rigid crank having length $l_1=0.05$ \si{m}. The slider mechanism is connected to the crank via a rigid connecting rod with approximate length $l_2=0.29$ \si{m}. The rotary motion of the motor is measured by an incremental encoder (8192 CPR). A compression spring is added to invoke additional nonlinear disturbances in the external force $F$ acting on the slider. The spring force characteristic as a function of displacement $d$ is given in Fig. \ref{fig:springLaw}. This characteristic was empirically determined for validation purposes later on. 
Experimental data are collected from the slider-crank setup for different torque signal inputs. These torque profiles, see Fig. \ref{fig:torque}, are chosen \textit{ad-hoc} to have diverse excitations of the system. These torque inputs are each time applied starting from two distinct motor positions, leading to a total of $S=20$ different trajectories of rotational movements $\theta$ and $\omega$. The measurements are sampled at 2000 Hz, resulting in $L=800$ samples per trajectory.
\vspace{-15pt}	
\begin{figure}[h]
	\centering
	\begin{subfigure}[H]{.47\columnwidth}
		\includegraphics[trim=6.5cm 11.9cm 7cm 12.1cm,clip=true,width=\columnwidth]{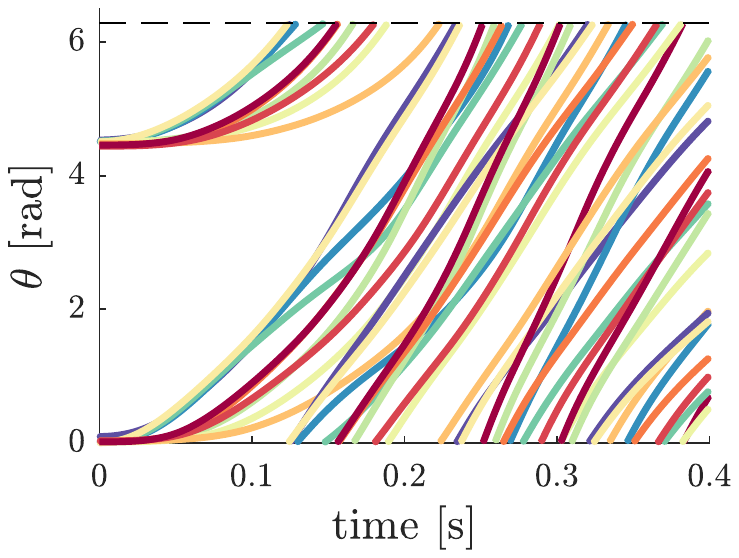}
	\end{subfigure}
	\begin{subfigure}[H]{.47\columnwidth}
		\includegraphics[trim=6.5cm 11.9cm 7cm 12.1cm,clip=true,width=\columnwidth]{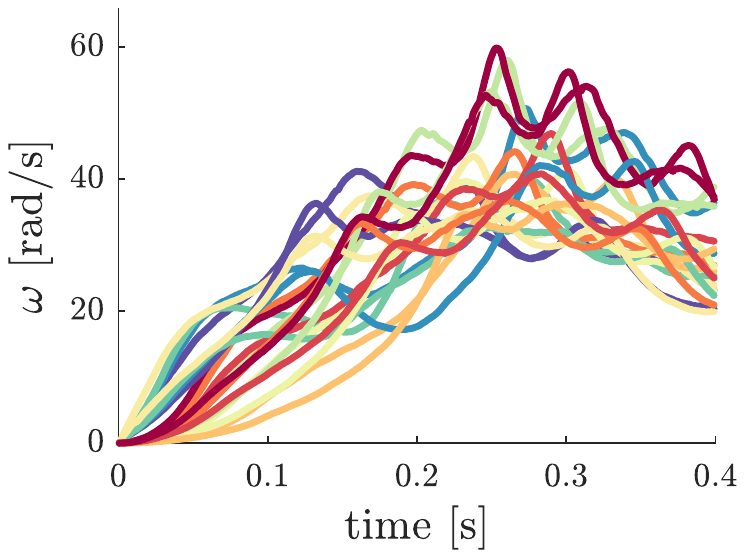}
	\end{subfigure}
	\vspace{-5pt}	
	\caption{Collected data from the slider-crank setup. }
	\label{fig:Experiments}
	\vspace{-10pt}
\end{figure}

\section{Neural Network Augmented Physics Model}\label{sec:Meth}
Having accurate predictive models of systems, such as the slider-crank setup presented in Section \ref{sec:SCsystem}, can advance the design and motion control of various manufacturing and industrial systems. To accommodate for partially unknown dynamics (unknown nonlinear disturbance $\mathcal{P}$ and uncertain physical parameters $\vectorstyle{p}$) arising in servo systems, the physics-based ordinary differential equations (ODE) are complemented with additional data-driven models. The physics-based derivative function $\vectorstyle{f}$ defines the behavior of the system state $\vectorstyle{x}$ for given control input $\vectorstyle{u}$. The known physical laws are encapsulated within $\vectorstyle{f}$ and characterized by physical parameters $\vectorstyle{p}$. The lack of knowledge about the unknown phenomena $\mathcal{P}$ is compensated for by introducing an additional input $\vectorstyle{z}$. 
\begin{equation}
\label{eq: phys_general}
\dot{\vectorstyle{x}}=\vectorstyle{f}\left( \vectorstyle{x}, \vectorstyle{u},\vectorstyle{z}; \vectorstyle{p}\right)
\end{equation}
The NNAP model here includes an artificial neural network $\vectorstyle{\eta}$ with the weights and biases of the neural nodes as learnable parameters $\vectorstyle{\alpha}$ to compensate for these unknown phenomena $\mathcal{P}$. The state measurements $\vectorstyle{x}$ and control input $\vectorstyle{u}$ are used to simultaneously update the neural network variables $\vectorstyle{\alpha}$ and physical parameters $\vectorstyle{p}$ to match the actual physical dynamical system behavior. Note that here, contrary to \cite{punjani2015deep}, the data-driven model is not learned on the basis of (additional) measurement data that relate directly to the unknown phenomena $\mathcal{P}$. A feedforward and a recurrent physics-based neural network model formalism are presented in the two next subsections, respectively. Subsequently, the optimization process is explained, followed by the implementation details on the slider-crank mechanism.

\subsection{Feedforward NNAP Model}
A feedforward model predicts the next state $\vectorstyle{x}_{k+1}$ for given control input $\vectorstyle{u}_k$ and current state $\vectorstyle{x}_k$. By repeatedly feeding the predicted state back as input during the subsequent time instance, the model becomes suitable for multistep predictions.
\begin{figure}[h]
	\vspace{0pt}
	\centering
	\includegraphics[trim=4.5cm 6.8cm 12cm  7.5cm,clip=true,width=0.85\columnwidth]{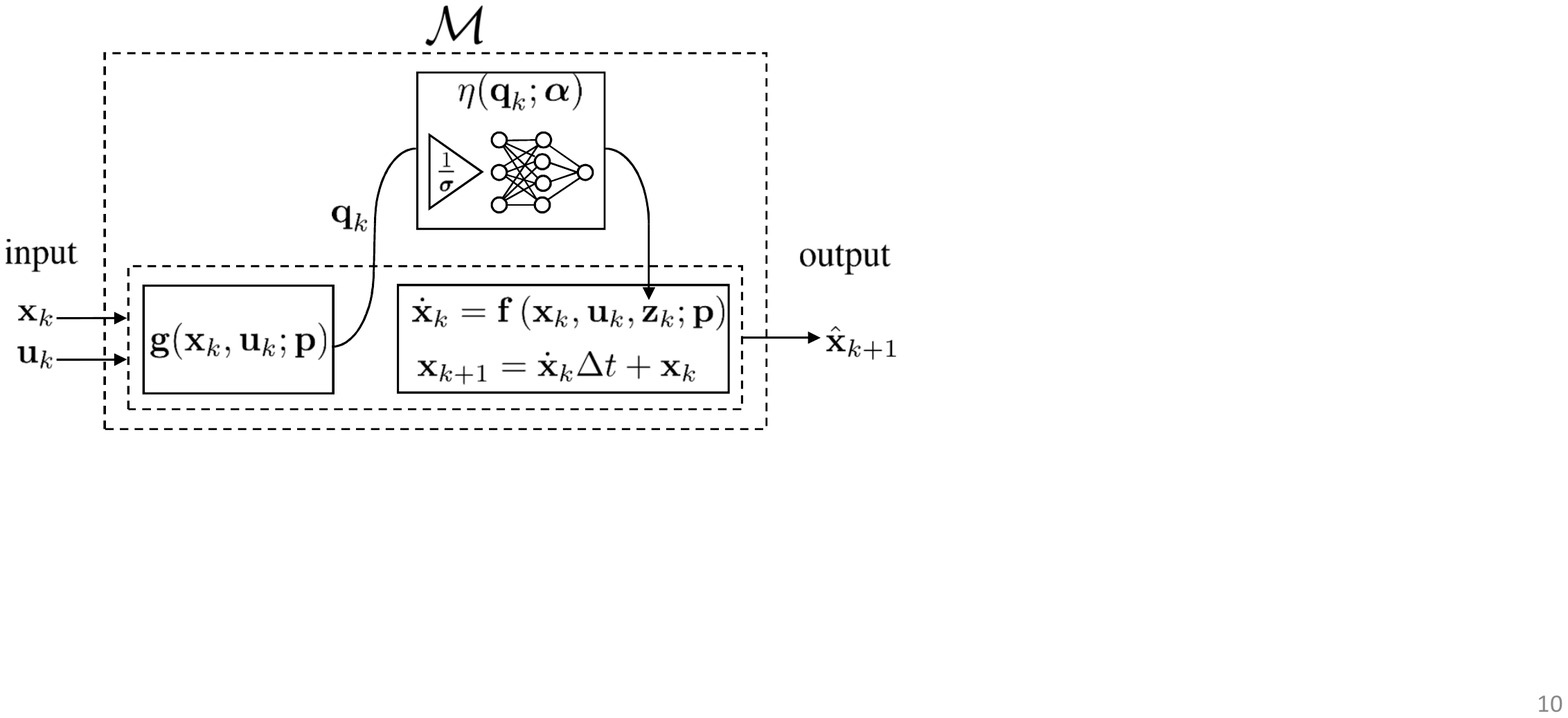}
	\caption{Feedforward NNAP modeling formalism.}
	\label{fig:hybrid}
	\vspace{0pt}
\end{figure}
The universal approximation theorem states that feedforward neural networks with one hidden layer can approximate any continuous function for inputs within a specific range \cite{hornik1989multilayer}. More specifically, we define $\vectorstyle{\eta}$, a one-hidden-layer rectified linear unit (ReLU) network of $n_h$ hidden units predicting a $n_o$ dimensional output. A ReLU model is chosen due to its demonstrated usefulness in various regression tasks \cite{lecun2015deep, ramachandran2017searching}. The $n_i$ dimensional input $\vectorstyle{q} \in \mathbb{R}^{n_i}$ to the neural network is defined by a physics-inspired network layer $\vectorstyle{g}$.
\begin{equation}
\label{eq:g}
\vectorstyle{q}=\vectorstyle{g}(\vectorstyle{x},\vectorstyle{u};\vectorstyle{p})
\end{equation}
The input vector $\vectorstyle{q}$ is scaled by the standard deviation ($\sigma_j$, $j=1,\cdots,n_i$) of the corresponding input elements to eliminate magnitude differences and consequently avoid dominance of particular dimensions. The corresponding transformation matrix $\Sigma=\text{diag} \left( \sigma_1^{-1}, \ldots, \sigma_{n_i}^{-1} \right)$ is determined in advance directly from the measurement data. Subsequently, the input undergoes a linear transformation by the weights $\matrixstyle{W}_h\in \mathbb{R}^{n_i\times n_h}$ of the hidden layer. The hidden unit activation function $\Upsilon$ will, after adding the bias vector $\vectorstyle{b}_h \in \mathbb{R}^{n_h}$, include the required nonlinearity in the function $\vectorstyle{\eta}$. Subsequently, the output is obtained via a linear output layer, defined by the weights $\matrixstyle{W}_o \in \mathbb{R}^{n_h\times n_o}$ and $\vectorstyle{b}_o\in \mathbb{R}^{n_o}$. We denote the ReLU model as being
\begin{equation}
\begin{aligned}
\vectorstyle{\eta}(\vectorstyle{q};\vectorstyle{\alpha})&= \matrixstyle{W}_o^T\Upsilon \left( \matrixstyle{W}_h^T \Sigma\vectorstyle{q}+ \vectorstyle{b}_h\right) + \vectorstyle{b}_o \\
\Upsilon(\cdot)&=\max\left(0,\cdot \right)
\end{aligned}
\end{equation}
The architecture of the feedforward NNAP models is given in Fig. \ref{fig:hybrid}. The output of the ReLU network $\vectorstyle{\eta}$, initialized by random weights and biases included in $\vectorstyle{\alpha}$, replaces the variable $\vectorstyle{z}$ in the ODE layer (\ref{eq: phys_general}). Consequently, the assumption is that $\vectorstyle{q}$  contains sufficient input information to estimate $\vectorstyle{z}$, by approximating $\mathcal{P}$ via a neural network $\vectorstyle{\eta}$. We distill the following derivative function at each time instant $k$.
\begin{equation}
\dot{\vectorstyle{x}}_k= \vectorstyle{f}\left( \vectorstyle{x}_k, \vectorstyle{u}_k,\vectorstyle{\eta}(\vectorstyle{q}_k;\vectorstyle{\alpha}); \vectorstyle{p}\right) \\
\end{equation} 
The derivative function now has a hybrid nature and is further used to propagate to the next state via Euler's method by choosing $\Delta t$ equal to the fixed sampling time of the measurement data. The combination of interconnected layers adds up to an overall feedforward network function $\mathcal{M}\left( \vectorstyle{x}_k, \vectorstyle{u}_k; \vectorstyle{p},\vectorstyle {\alpha}\right)$ that estimates the state $\vectorstyle{x_{k+1}}$ given the prior state $\vectorstyle{x_{k}}$ and input $\vectorstyle{u_{k}}$.
\begin{equation}
\hat{\vectorstyle{x}}_{k+1}=\mathcal{M}\left( \vectorstyle{x}_k, \vectorstyle{u}_k; \vectorstyle{p},\vectorstyle {\alpha}\right) 
\label{eq:hybrid}
\end{equation}
The feedforward NNAP model $\mathcal{M}$ predicts only the state of the subsequent timestep. The model $\mathcal{M}$ encompasses a static relation that can describe dynamical behavior once prior state predictions are fed back as model input.
\subsection{Recurrent NNAP model}
\begin{figure}[h]
	\includegraphics[trim=8.8cm 9.5cm 8cm 10cm,clip=true,angle=-90,width=0.9\columnwidth]{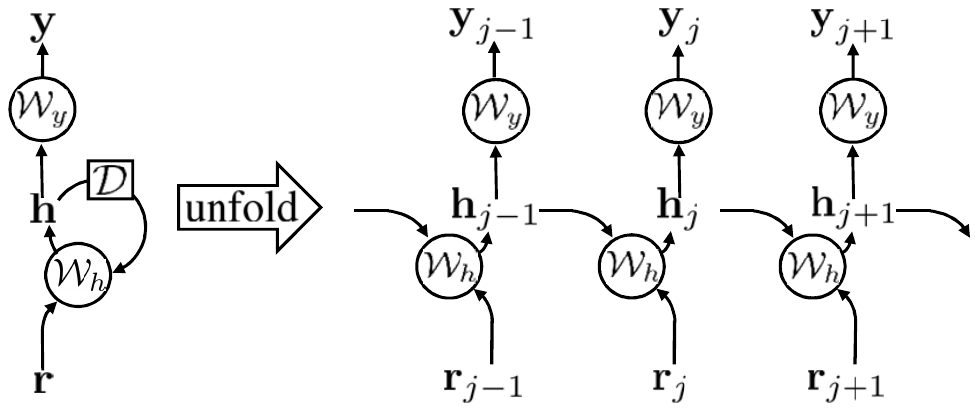}
	\caption{Recurrent neural network.}
	\label{fig:RNNunfold}
\end{figure}

\begin{figure}[h]
	\vspace{-15pt}
	\includegraphics[trim=7.6cm 9.8cm 9.5cm 9.9cm,clip=true,angle=-90,width=0.9\columnwidth]{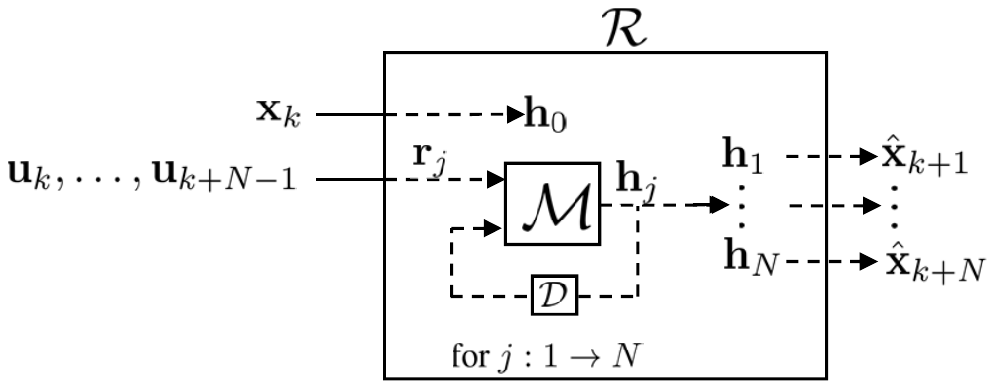}
	\caption{Recurrent NNAP modeling formalism.}
	\label{fig:hybridRNN}
	\vspace{-5pt}
\end{figure}

Recurrent neural networks (RNN) can cope with timeseries due to their ingrained dynamic nature \cite{pascanu2013difficulty,lecun2015deep}. These models are penalized for estimation errors over a larger time horizon, aiming for improved multistep prediction capabilities. Fig. \ref{fig:RNNunfold} illustrates a schematic overview of a basic RNN. The operator $\mathcal{D}$ induces a one-step time delay of the internal state $\vectorstyle{h}$. If we unfold the RNN in time, we obtain a computational graph that contains shared weights $\vectorstyle{\psi}$ at each time instance $j$. The total prediction error is an accumulation of the error at each timestep. The gradients of this virtual multilayer network are required to update $\vectorstyle{\psi}$ and can be derived via backpropagation through time (BPTT), which calculates the gradients based on the chain rule of differentiation \cite{pascanu2013difficulty}. 
A RNN model accepts an input sequence of $N$ steps by processing each input element $\vectorstyle{r}_j$ for $j \in \{1,\ldots,N\}$  at a time. These models contain information about prior inputs by updating an internal state vector $\vectorstyle{h}_j$ via a (nonlinear) function $\mathcal{W}_h$.  Thereafter, an output relation $\mathcal{W}_y$ maps the internal state to the model output $\vectorstyle{y}$. The functions $\mathcal{W}_h$ and $\mathcal{W}_y$ typically contain network parameters $\vectorstyle{\psi}$ that are optimized via gradient-based algorithms.
\begin{equation}
\begin{aligned}
\vectorstyle{h}_j&=\mathcal{W}_h(\vectorstyle{r}_j,\vectorstyle{h}_{j-1}; \vectorstyle{\psi})\\
\vectorstyle{y}_j&=\mathcal{W}_y(\vectorstyle{h}_j;\vectorstyle{\psi})
\label{eq:unf}
\end{aligned}
\end{equation}
A recurrent NNAP model $\mathcal{R}$ is constructed to predict a state trajectory $\{\vectorstyle{x}_{k+1},\ldots, \vectorstyle{x}_{k+N}\}$ for given input sequence $\{ \vectorstyle{u}_k, \ldots ,\vectorstyle{u}_{k+N-1}\}$, starting from state $\vectorstyle{x}_k$. The dimension of the internal state vector $\vectorstyle{h}$ is typically a hyperparameter that needs to be tuned depending on the complexity of the problem. Here, we assume that the dimension of the state of $\vectorstyle{h}$ equals the dimension of the state vector $\vectorstyle{x}$. The corresponding initial value $\vectorstyle{h}_0$, typically initialized as zero, is chosen to equal $\vectorstyle{x_k}$.
This allows to use the feedforward NNAP function $\mathcal{M}$, detailed in Fig. \ref{fig:hybrid}, as an update function $\mathcal{W}_h$ of the internal state $\vectorstyle{h}$ in (\ref{eq:unf}). This forces the hidden state $\vectorstyle{h}$ to mimic the behavior of the system state $\vectorstyle{x}$. Consequently, the output function $\mathcal{W}_y$ in (\ref{eq:unf}) includes an identity transformation because we can consider $\vectorstyle{y}_j=\vectorstyle{h}_j=\vectorstyle{x}_{k+j} \text{ for }j \in \{1,\ldots,N\}$. Fig. \ref{fig:hybridRNN} illustrates the architecture of $\mathcal{R}$. The identification of the recurrent NNAP model $\mathcal{R}$ parameters $\vectorstyle{\psi}$ comes down to optimizing the variables $\vectorstyle{p}$ and $\vectorstyle{\alpha}$ included in the feedforward NNAP model $\mathcal{M}$ in (\ref{eq:hybrid}).
\begin{equation}
\begin{bmatrix}
\hat{\vectorstyle{x}}_{k+1},\ldots, \hat{\vectorstyle{x}}_{k+N}
\end{bmatrix}=
\mathcal{R}(\vectorstyle{x}_k, \vectorstyle{u}_k, \ldots ,\vectorstyle{u}_{k+N-1}; \vectorstyle{p},\vectorstyle{\alpha})
\label{eq:hybridRNN}
\end{equation}
\subsection{Optimization}
The identification of the NNAP model requires the simultaneous optimization of the neural ($\vectorstyle{\alpha}$) and physical ($\vectorstyle{p}$) parameters by minimizing the cost function $\mathcal{L}$. 
\begin{equation}
\label{eq:cost}
\mathcal{L}(\vectorstyle{\alpha},\vectorstyle{p}) = \frac{1}{S} \frac{1}{L} \frac{1}{N}\sum_{s=1}^{S} \sum_{i=0}^{L-N} \sum_{k=i+1}^{i+N}  (\vectorstyle{x}^s_k-\hat{\vectorstyle{x}}^s_k)^T \matrixstyle{C}  (\vectorstyle{x}^s_k-\hat{\vectorstyle{x}}^s_k)
\end{equation}
The function $\mathcal{L}$ penalizes the discrepancies between the measured ($\vectorstyle{x}$) and predicted ($\hat{\vectorstyle{x}}$) states, scaled by a diagonal matrix $\matrixstyle{C}$ to accommodate for magnitude differences between the state variables. The loss function $\mathcal{L}$ incorporates $S$ trajectory signals, for which each signal contains $L$ samples of the state $\vectorstyle{x}$ and control input $\vectorstyle{u}$. 
\begin{figure}[h]
	\centering
	\begin{subfigure}[]{.355\columnwidth}
		\caption{training data $\mathcal{M}$}
		\vspace{-2pt}
		\label{fig:train_FFNN}
		\includegraphics[trim=4.6cm 3.9cm 9.1cm 20.6cm,clip=true,angle=-90,width=\columnwidth]{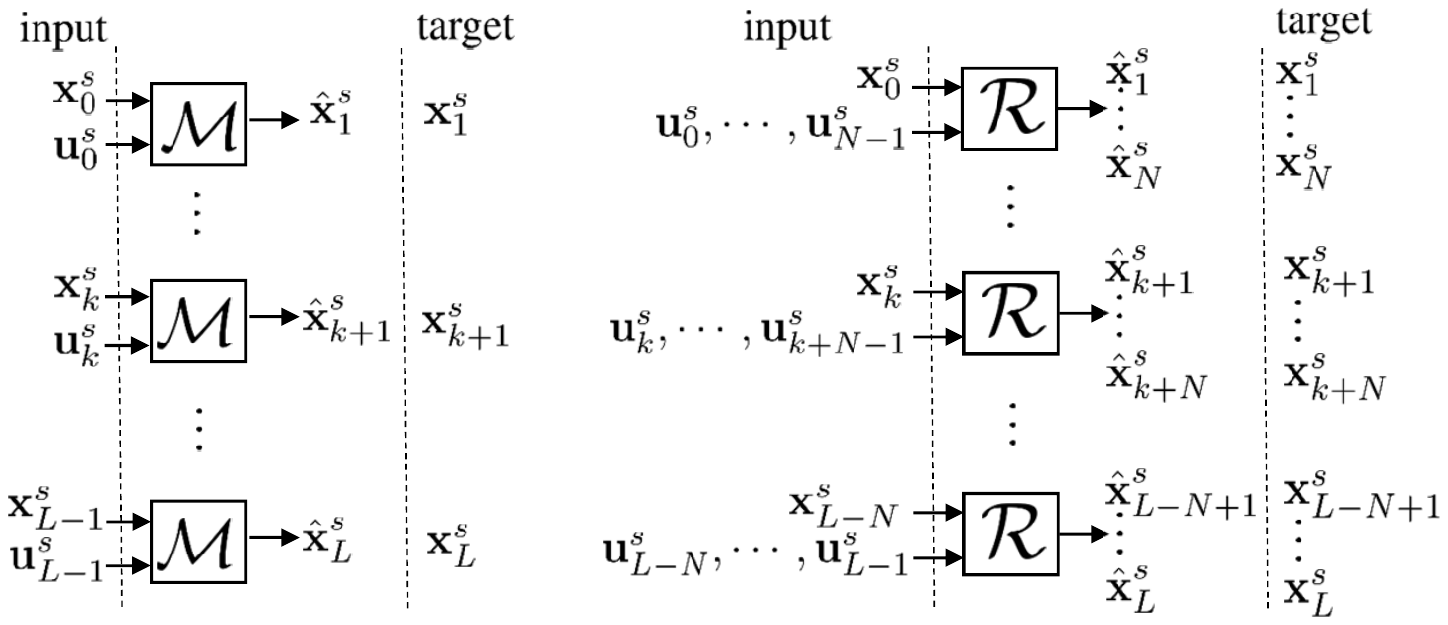}
	\end{subfigure}
	\hspace{0.15 cm}
	\begin{subfigure}[]{.585\columnwidth}
		\caption{training data $\mathcal{R}$}
		\vspace{-2pt}
		\label{fig:train_RNN}
		\includegraphics[trim=4.6cm 10cm 9.1cm 11.1cm,angle=-90,clip=true,width=\columnwidth]{FIG11}
	\end{subfigure}
	\vspace{-20pt}
	\caption{Schematic overview of training data specifications for trajectory $s\in \{1, \ldots,S\}$ existing of $L$ training samples.}
	\label{fig:trainTest}
\end{figure}
Figure \ref{fig:train_FFNN} illustrates the evaluation of each trajectory sample by $\mathcal{M}$, to match the subsequent state value (we have here $N=1$). The recurrent model $\mathcal{R}$ is constructed on the basis of an initial state value and control sequence of $N$ steps, as depicted in Fig. \ref{fig:train_RNN}. The derivation of the analytical gradients $\nabla_{p}\mathcal{L}$ and $\nabla_{\alpha}\mathcal{L}$ imposes that, apart from the neural network $\vectorstyle{\eta}$, the physics layers also need to be differentiable. The gradient information is used to simultaneously update $\vectorstyle{\alpha}$ and $\vectorstyle{p}$ based on a gradient descent approach with step size $\delta$.   
\begin{equation}
\label{eq:gradient}
\begin{aligned}
\vectorstyle{p}_{i+1} = \vectorstyle{p}_{i}-\delta_i\nabla _p \mathcal{L} \\
\vectorstyle{\alpha}_{i+1} = \vectorstyle{\alpha}_{i}-\delta_i\nabla _\alpha \mathcal{L}
\end{aligned}
\end{equation} 
Note that the difference between the feedforward model $\mathcal{M}$ and recurrent architecture $\mathcal{R}$ lies in the different training process as the gradients of the latter are derived by BPTT to allow robust predictions of longer time series. Nevertheless, once the recurrent model $\mathcal{R}$ converges, it can be reworked to a more simple feedforward structure $\mathcal{M}$ because both architectures ingrain the same parameters $\vectorstyle{p}$ and $\vectorstyle{\alpha}$.

\subsection{Implementation}
\begin{figure}[b]
	\vspace{-1 cm}
	\centering
	\includegraphics[trim=0cm 0cm 0cm  3.3cm,clip=true,angle=-90,width=1\columnwidth]{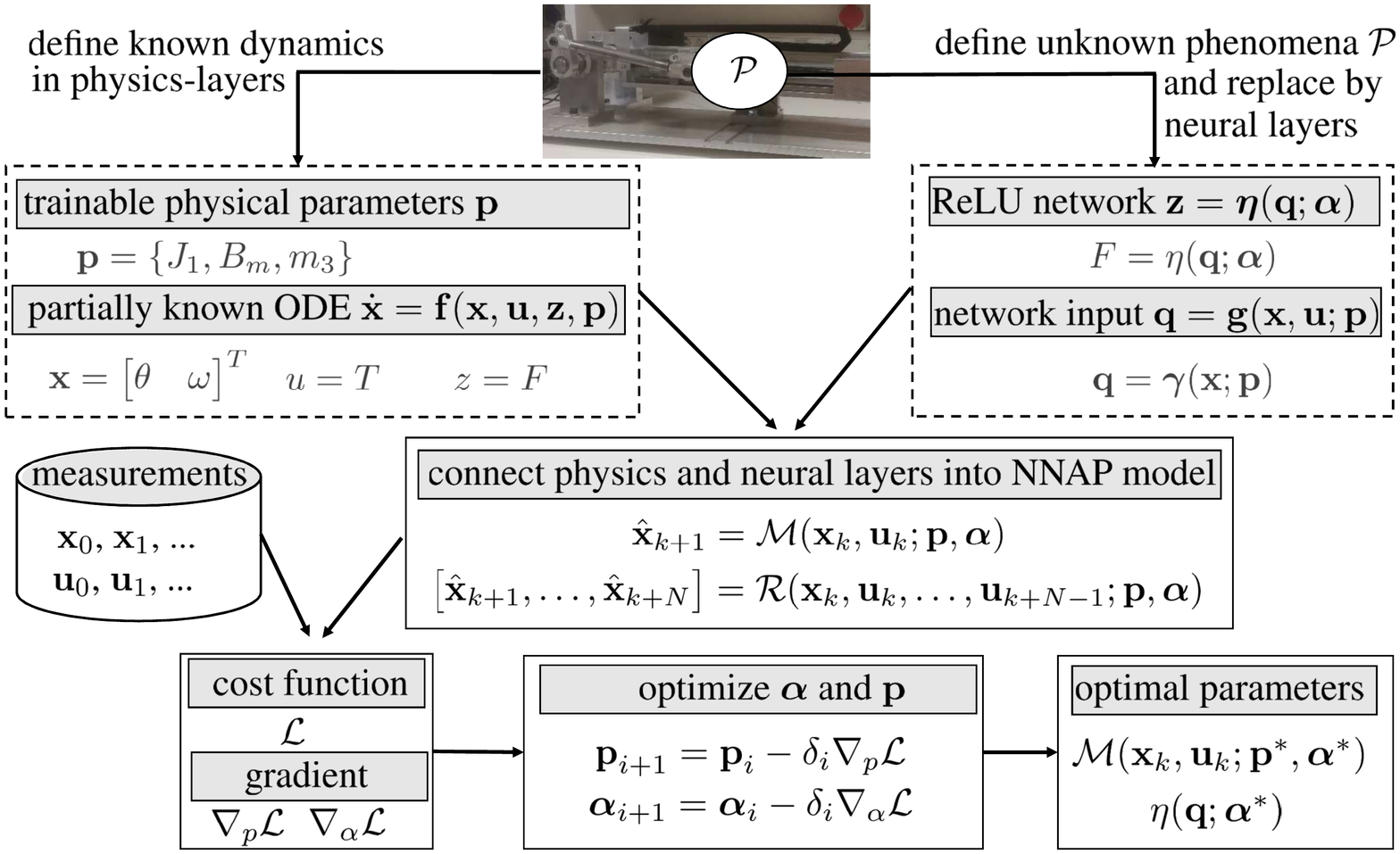}
	\vspace{-10pt}
	\caption{Practical implementation of the NNAP model.}
	\label{fig:implementation}
	\vspace{-10pt}
\end{figure}

The presented NNAP modeling formalism is applied on the slider-crank mechanism detailed in Section \ref{sec:SCsystem}. The general workflow is illustrated in Fig. \ref{fig:implementation}. First, the known dynamics of the system are distilled and described by their physical laws. As mentioned before, the dynamics of the slider-crank mechanism are known, except for the state-dependent load interactions $\mathcal{P}$ acting on the slider. The incomplete physics-based relations are combined into an ODE $\vectorstyle{f}$ that accepts an additional input $z=F$, as explained in Appendix \ref{App:Model}. Subsequently, the uncertain physical parameters $\vectorstyle{p}$ are defined. In the slider-crank case, we consider the parameter set $\vectorstyle{p}= \{J_1, B_m, m_3 \}$ as unknown. Other possibilities are discussed in Section \ref{sec:Simultaneous}.

Next, the unknown load interactions $\mathcal{P}$ are replaced by a randomly initialized ReLU network $\eta$. This predictor of the load force $F$ contains one hidden layer of $n_h=32$ hidden units, determined via hyperparameter tuning. The network input $\vectorstyle{q}$ needs to provide sufficient information to the network $\eta$. Therefore, expert knowledge is used to define $\vectorstyle{q} = \begin{bmatrix}  d & v \end{bmatrix}^T$ by choosing the input function $\vectorstyle{g}$ equal to the geometrical relations $\vectorstyle{\gamma}$ in (\ref{eq:geometrical}). Other options for $\vectorstyle{g}$ and their effect are discussed in Section \ref{seq:q}. The relations defined in the functions $\vectorstyle{g}(.,\vectorstyle{p})$, $\eta(.,\vectorstyle{\alpha})$ and  $\vectorstyle{f}(.,\vectorstyle{p})$ are implemented as successive custom network layers, that result into the overall NNAP model $\mathcal{M}(., \vectorstyle{p},\vectorstyle{\alpha})$, as shown in Fig. \ref{fig:hybrid}. Moreover, this feedforward model can be implemented as a recurrent NNAP model $\mathcal{R}(., \vectorstyle{p},\vectorstyle{\alpha})$  by adding the feedback loops shown in Fig. \ref{fig:hybridRNN}. In practice, these network architectures are implemented by using the Keras API \cite{chollet2015keras} with TensorFlow \cite{abadi2016tensorflow} backend in Python. The physical laws are implemented as custom (differentiable) network layers for which the parameters $\vectorstyle{p}$ are defined as trainable variables. This library employs automatic differentiation, which provides the analytical expressions for the gradients $\nabla_{p}\mathcal{L}$ and $\nabla_{\alpha}\mathcal{L}$. The availability of analytical gradient information enables a joint identification of both $\vectorstyle{p}$ and $\vectorstyle{\alpha}$ by using a gradient based Adam optimizer that processes the data in mini-batches of 200 samples \cite{kingma2014adam}.

\section{Results and Discussion} \label{sec:Results}

The objective of the joint optimization of $\vectorstyle{p}$ and $\vectorstyle{\alpha}$ is twofold. First, the prediction performances of the optimized NNAP model are assessed by analyzing the root mean squared error (RMSE) of the trajectory prediction of the key variable $\omega$. The presented results are validated via leave-one-out-cross-validation (LOOCV). This implies that the model is trained by $S=19$ signals and validated by the trajectory excluded from
the training set. This is repeated for all 20 signals in order to
develop a fair performance benchmark. Unless otherwise
stated, we will perform the experiments on the feedforward
architecture $\mathcal{M}$ because inherently, both architectures $\mathcal{M}$ and
$\mathcal{R}$ include the same parameters to be identified. Secondly, the convergence and identifiability of the physical parameters $\vectorstyle{p}$ and the neural network $\eta$ will be analyzed to validate if new physical insights can be gained about the physical parameters and the unknown load interactions captured by $\eta$.
\subsection{Influence of the Neural Network Inputs}
\label{seq:q}
\begin{figure}[h]
	\centering
	\begin{subfigure}[H]{.39\columnwidth}
		\caption{prediction RMSE}
		\includegraphics[trim=7cm 10.9cm 7.5cm 11.4cm,clip=true,width=\columnwidth]{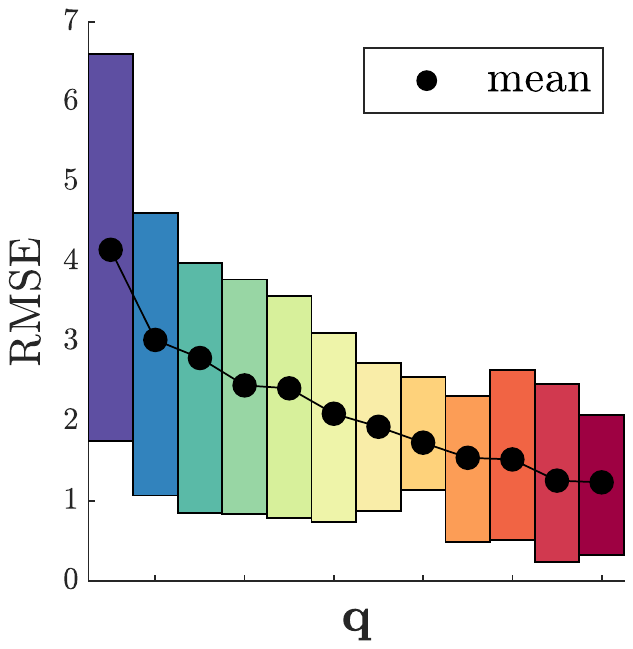}
		\label{fig:rmse}
	\end{subfigure}
	\begin{subfigure}[H]{.59\columnwidth}
		\caption{multistep predictions of rot. velocity}
		\includegraphics[trim=5.9cm 11.3cm 6.3cm 11.7cm,clip=true,width=\columnwidth]{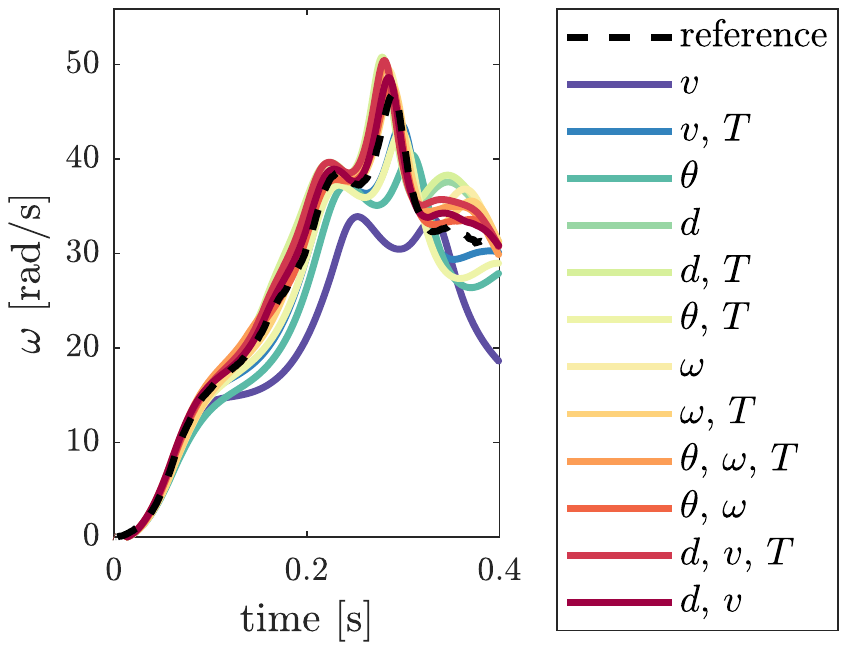}
		\label{fig:inputTS}
	\end{subfigure}
	\vspace{-10pt}
	\caption{Influence of the input information included in $\vectorstyle{q}$. The left plot represents 90\% of the most accurate results obtained on test data via LOOCV. A prediction of test signal $T_9$ is illustrated on the right. }
	\vspace{-10pt}
	\label{fig:Inputinfluence}
\end{figure}

The prediction performance of the NNAP model $\mathcal{M}$ highly relies on the information that can be captured by the neural network $\eta$ during the joint optimization of the parameters in $\vectorstyle{\alpha}$ and $\vectorstyle{p}$. The ability of $\eta$ to compensate for the unmodeled physical relations in $\vectorstyle{f}$ depends on the input $\vectorstyle{q}$, defined by the function  $\vectorstyle{g}$. For this case, expert knowledge is used to define $\vectorstyle{g}$ as the geometrical relations  $\vectorstyle{\gamma}$ in (\ref{eq:geometrical}), to provide $\vectorstyle{q}=\begin{bmatrix} d & v \end{bmatrix}^T$ as input to $\eta$. However, in practice, the dependencies of the unknown phenomena $\mathcal{P}$ are not known. Therefore, the required information to the neural network can be determined by analyzing the obtained prediction accuracy of $\mathcal{M}$ for varying function candidates of $\vectorstyle{g}$. Variables such as $\{\theta$, $\omega\}$ and $\{T\}$ can be obtained by choosing $\vectorstyle{g}$ equal to an identity transformation because they are directly included in the state $\vectorstyle{x}$ or control input $u$, respectively. Figure \ref{fig:inputTS} illustrates the time evolution of the angular velocity $\omega$ on the basis of the discrete model $\mathcal{M}$ for varying  inputs $\vectorstyle{q}$. We perform a multistep prediction as the state  $\hat{\vectorstyle{x}}_{k+1}$ in (\ref{eq:hybrid}) is the subsequent state input $\vectorstyle{x}_{k}$ to $\mathcal{M}$ in the next time instant. Errors are propagated each time when $\mathcal{M}$ is evaluated, explaining the diverging trajectories with respect to the reference signal. The lowest average RMSE is obtained for $\vectorstyle{q}=\begin{bmatrix} d & v \end{bmatrix}^T$. This insight makes sense knowing that the spring load and possible additional friction phenomena, both contained in the overall force $F$, are not included in the physics equations and, thus, should be approximated by the network $\eta$. The set of experiments for which the required input information (i.e. set $\{d,v\}$  or the equivalently $\{\theta,\omega \}$) is not fed to the neural network $\eta$ clearly lead to inferior trajectory prediction accuracies. This experiment revealed the importance of the input information to the neural network $\eta$, at the same time showing that the NNAP model $\mathcal{M}$ converges if indeed sufficient information is fed to $\eta$.

\subsection{Region of Operations}
\begin{figure}[b]
	\centering
	\includegraphics[trim=4.4cm 11.4cm 4.7cm 11.6cm,clip=true,width=1\columnwidth]{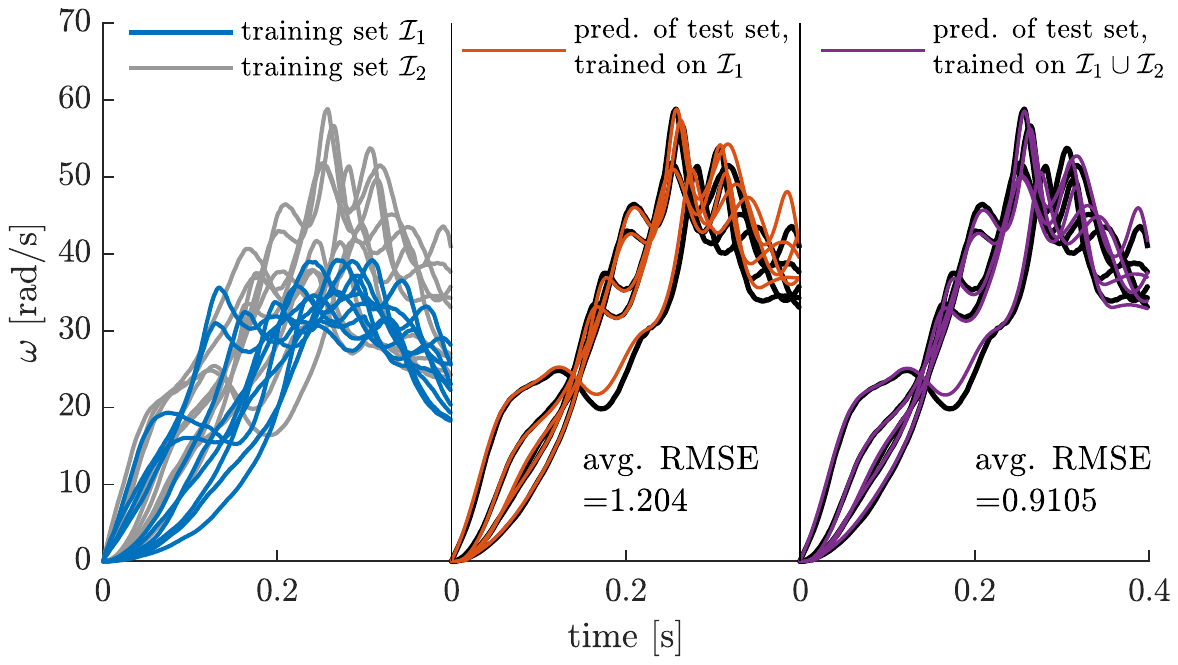}
	\vspace{-10pt}
	\caption{Influence of having training data in the neighborhood of test signals. The left plot indicates the subdivision of the training set. The central and right plots show the prediction result on test data of a model trained on halved (i.e. limited region of operation) and full training set respectively.}
	\label{fig:RoO}
\end{figure}
Neural networks can capture complex relations. However, as is generally known, in their current form, they fail to extrapolate well outside their training region. Physics-based relations, on the other hand, can cover a wide range of operating points because they are substantiated by general physical laws. By using physics-inspired custom layers, we alleviate the modeling complexity a neural network of the complete system needs to capture. To validate the robustness and stability of the overall NNAP model, we define the set $\mathcal{I}_1$ that includes 10 trajectories with low operating speed. The central plot in Fig. \ref{fig:RoO} shows the predictions of the signals operating at the highest speed, by training the NNAP model only with trajectories included in $\mathcal{I}_1$. The right plot in Fig. \ref{fig:RoO} illustrates the prediction performances obtained by a LOOCV experiment, covering the entire region of operations in the training set. As expected, the average RMSE indicates that the prediction accuracy deteriorates once the test signals exceed the training region. The NNAP model, however, still follows the general dynamic behavior outside the training region. These results indicate the benefit of having a solid physics-inspired basis within the dynamic NNAP model.

\subsection{Recurrent NNAP Model Multistep Prediction}
\begin{figure}[b]
 	\vspace{-0.5cm}
	\centering
	\begin{subfigure}[H]{.49\columnwidth}
		\caption{accuracy}
		\label{fig:RNNsteps1}
		\includegraphics[trim=6.5cm 11.4cm 6.9cm 11.5cm,clip=true,width=\columnwidth]{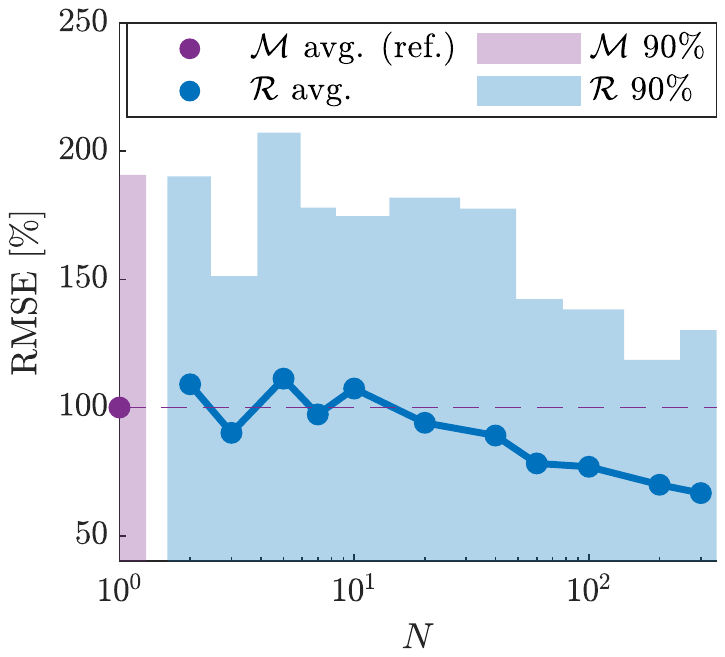}
	\end{subfigure}
	\begin{subfigure}[H]{.49\columnwidth}
		\caption{training time per epoch}
		\label{fig:timeRNN}
		\includegraphics[trim=6.5cm 11.4cm 7cm 11.5cm,clip=true,width=\columnwidth]{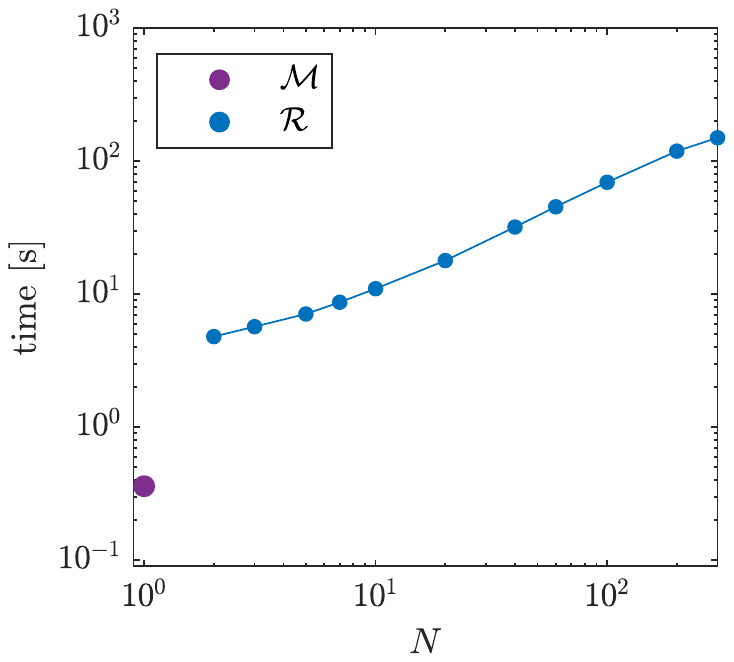}
	\end{subfigure}
	\caption{Influence of step size $N$ on the accuracy and required training time per epoch. The accuracy of the multistep predictions is tested by LOOCV. The left plot shows the average accuracy and the interval that includes 90\% of the most accurate test signals. The right plot illustrates the time required to train one epoch for different number of timesteps $N$ of the recurrent NNAP model $\mathcal{R}$.}
	\vspace{-10pt}
\end{figure}

\begin{figure}[h]
	\centering
	\begin{subfigure}[H]{.49\columnwidth}
		\caption{prediction}
		\label{fig:TS_RNN}
		\vspace{-13pt}
		\includegraphics[trim=6.5cm 11.4cm 7cm 11.5cm,clip=true,width=\columnwidth]{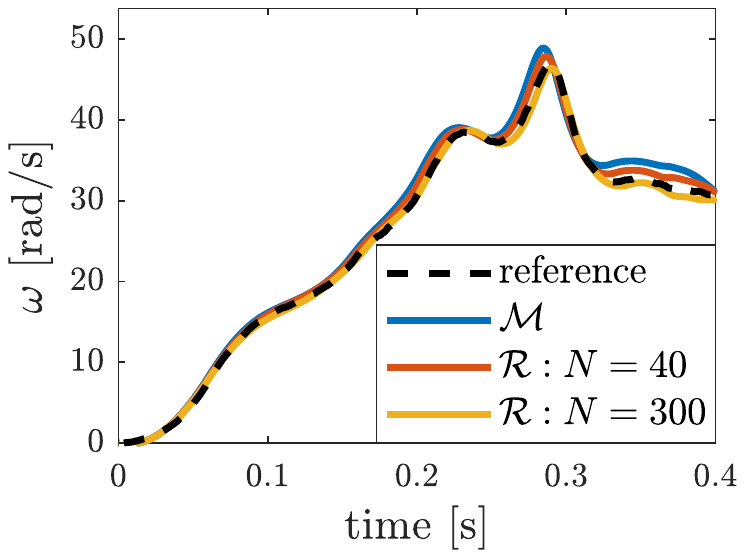}
	\end{subfigure}
	\begin{subfigure}[H]{.49\columnwidth}
		\caption{prediction error}
		\label{fig:TS_error_RNN}
		\vspace{-13pt}
		\includegraphics[trim=6.5cm 11.4cm 7cm 11.5cm,clip=true,width=\columnwidth]{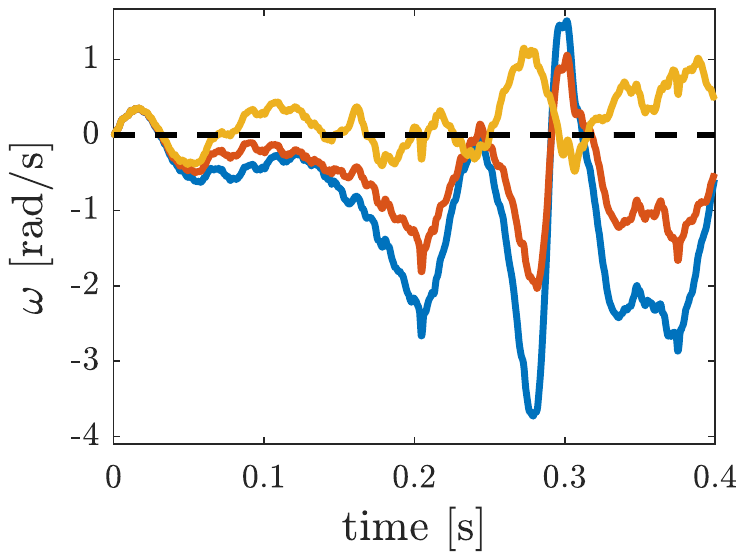}
	\end{subfigure}
	\vspace{-5pt}
	\caption{Multistep prediction of test signal $T_9$. }
	\label{fig:TS_RNN_analysis}
\end{figure}

The training process of a feedforward NNAP model $\mathcal{M}$ only penalizes the prediction error of the next timestep. A recurrent  NNAP model $\mathcal{R}$, on the other hand, penalizes the average accuracy of a trajectory sample. Because the model learns from the general dynamics this way, multistep predictions can be potentially improved. The influence of the length of the trajectory sequence $N$ of the recurrent model $\mathcal{R}$ of the slider-crank system is depicted in Fig. \ref{fig:RNNsteps1}. The aforementioned converged model $\mathcal{M}$ is used to initialize $\mathcal{R}$. Each model $\mathcal{R}$ is trained for 300 epochs in order to check whether these recurrent structures can improve the prediction capabilities. The average accuracy of the feedforward model $\mathcal{M}$ after 300 additional training epochs serves as a reference. The LOOCV reveals that the recurrent  model $\mathcal{R}$, in general, does not perform better for unseen trajectories for small values of $N$. However, the accuracy on test data clearly outperforms the feedforward model $\mathcal{M}$ for training on larger time sequences $N$. The average RMSE of the test data was reduced by 33.4\% for a model with $N=300$ compared to the feedforward model $\mathcal{M}$. This improved prediction accuracy, however, comes with increased computational costs as recurrent neural networks virtually unfold into more complex multilayer networks in order to apply BPTT (as illustrated in Fig. \ref{fig:RNNunfold}). The required computational time (training time per epoch) almost follows a power-law relation w.r.t. the sequences length $N$ used during the training, as can be observed in Fig. \ref{fig:timeRNN}. This experiment was performed on a processing unit including a 6-core Intel i5-8400 CPU with 16 GB RAM.
 Fig. \ref{fig:TS_RNN} illustrates an example of a multistep prediction of a test data trajectory of $L=800$ time instances by a NNAP model. The recurrent neural network $\mathcal{R}$ indicates better prediction performances compared to $\mathcal{M}$. The higher the length $N$ of the time sequences used during training, the more the NNAP models are able to capture the global dynamics of the slider-crank mechanism. A FFNN architecture only incorporates the prediction accuracy of one timestep within the cost function (\ref{eq:cost}). Naturally, drift occurs because the prediction error at a specific time instance is propagated through the subsequent timestep.  A RNN structure accommodates for these drift phenomena by incorporating the prediction accuracy for an entire trajectory segment within the cost function. Therefore, the RNN model learns to capture the dynamics for a longer time horizon, which leads to more robust prediction capabilities that better track the measured trajectories, as shown in Fig. \ref{fig:TS_error_RNN}.

\vspace{-0.5cm}
\subsection{Identifiability of the Physical Parameters} \label{sec:Simultaneous}
\begin{figure}[t!]
	\centering
	\begin{subfigure}[H]{.49\columnwidth}
		\caption{\vectorstyle{p} = \{$m_3$,$J_1$,$B_m$\}}
		\label{fig:m3_opt}	
		\includegraphics[trim=7.4cm 9.2cm 7.8cm 9.8cm,clip=true,width=\columnwidth]{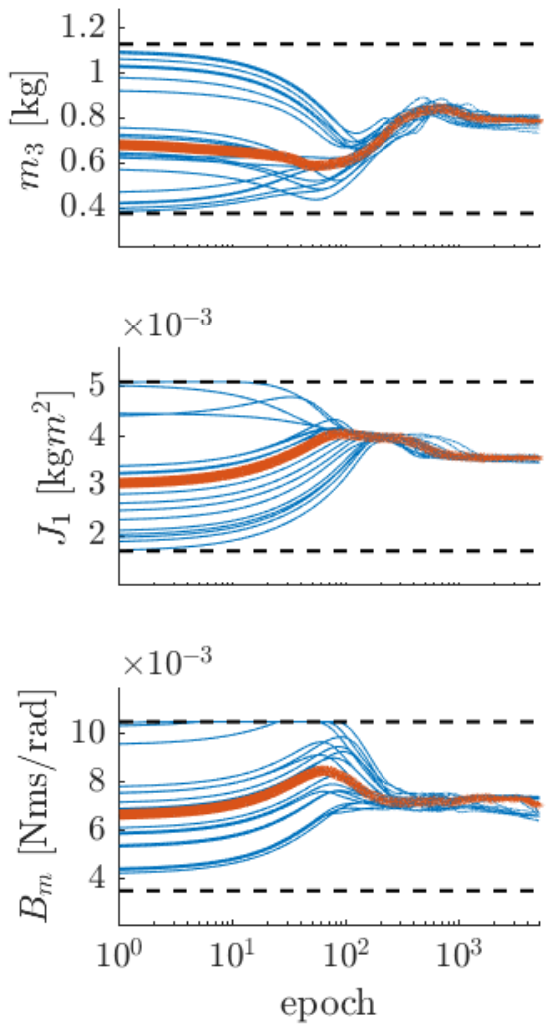}
	\end{subfigure}
	\begin{subfigure}[H]{.49\columnwidth}
		\caption{\vectorstyle{p} = \{$l_1$,$J_1$,$B_m$\}}
		\label{fig:l1_opt}
		\includegraphics[trim=7.4cm 9.2cm 7.8cm 9.8cm,clip=true,width=\columnwidth]{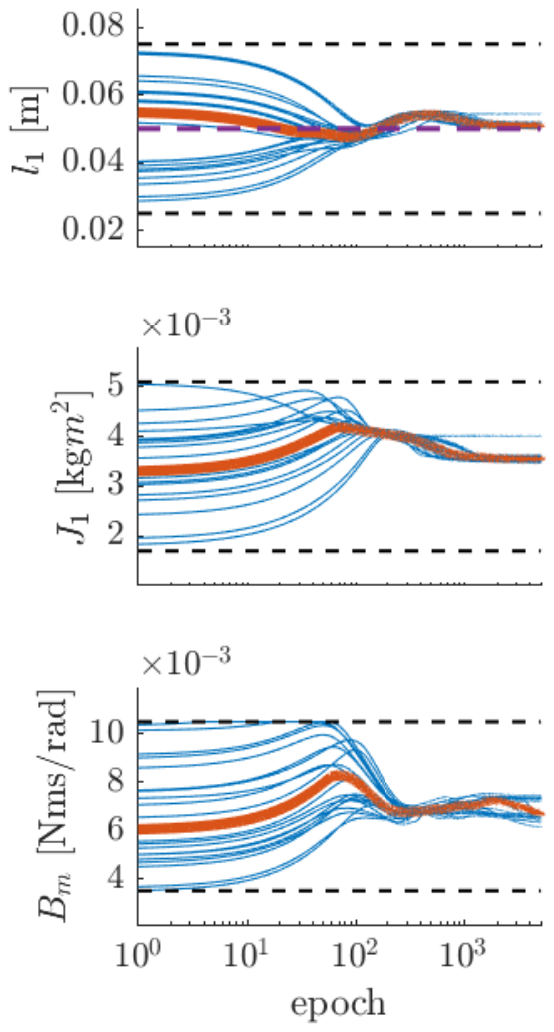}
	\end{subfigure}
	\caption{Convergence history of physical parameters $\vectorstyle{p}$ during simultaneous optimization of $\vectorstyle{p}$ and neural network parameters $\vectorstyle{\alpha}$. The left and right plots indicate the convergence history of an optimization process of parameters  $\vectorstyle{p} = \{m_3,J_1,B_m\}$ and $\vectorstyle{p} = \{l_1,J_1,B_m\}$, respectively.}
	\vspace{-10pt}
	\label{fig:physpar_3opt}
\end{figure}
Besides having an accurate NNAP model for multistep predictions, we aim to discover new physical insights. Because the physical parameters $\vectorstyle{p}$ are updated simultaneously with the identification of $\eta$, research should be devoted to assess the identifiability of these parameters. The physical layers, including the model described in Appendix \ref{App:Model}, contain 7 parameters $\vectorstyle{p}=\{m_1,m_2,m_3,l_1,l_2,J_1,B_m\}$. In practice, the ability to uniquely identify a parameter relates both to its sensitivity to the model output and its interference with other parameters during the optimization process. For the presented NNAP model in particular, one should also analyze whether the influence of the neural network $\eta$ is sufficiently uncorrelated with parameters $\vectorstyle{p}$ so that the neural network does not completely take over in the optimization process. Therefore, in Appendix \ref{App:SensitivityAnalysis}, a sensitivity analysis is applied to select the identifiable parameter sets, indicating that $l_1$ and $m_3$ need to be separated. Figure \ref{fig:physpar_3opt} illustrates the convergence history of parameter sets\footnotemark  $\vectorstyle{p}=\{m_3,J_1,B_m\}$ and $\vectorstyle{p}=\{l_1,J_1,B_m\}$ during the gradient-based optimization process of both $\vectorstyle{p}$ and $\eta$.
\footnotetext{Because these parameters are implemented as trainable weights within the custom physics layers, the optimization process requires parameter scaling to cope with the differences in order of magnitude between the variables in $\vectorstyle{p}$.} 
The initial guess is randomly chosen $\pm$ 50\% around the nominal value for each optimization run during the training procedure (LOOCV). The remaining physical parameters are held fixed during the optimization process. Nevertheless, we let them vary $\pm$ 10\% around their nominal value (see Table \ref{table:parReal}) in each experiment to indicate their insignificance when identifying the NNAP model. The orange line indicates the median, and the optimization boundaries are defined by the black dotted lines. The actual value of the geometrical parameter $l_1$ can easily be measured and is indicated by the purple line. The presented results indicate convergence of the proposed parameter sets. However, note that convergence cannot be obtained for all parameter combinations, as is further explained in Appendix \ref{App:SensitivityAnalysis}.

\subsection{Retrieving Physical Insights from the Neural Network}
\begin{figure}[h]
	\centering
	\begin{subfigure}[H]{.49\columnwidth}
		\caption{$\eta(\vectorstyle{q};\vectorstyle{\alpha}_0)$}
		\includegraphics[trim=6.5cm 11.9cm 6.6cm 12.1cm,clip=true,width=1\columnwidth]{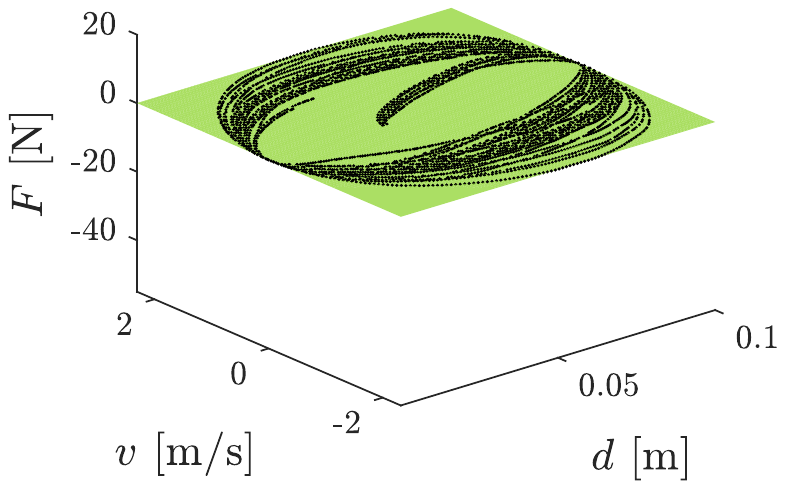}
		\label{fig:NNini_dv}
	\end{subfigure}
	\begin{subfigure}[H]{.49\columnwidth}
		\caption{$\eta(\vectorstyle{q};\vectorstyle{\alpha}^*)$}
		\includegraphics[trim=6.5cm 11.9cm 6.6cm 12.1cm,clip=true,width=1\columnwidth]{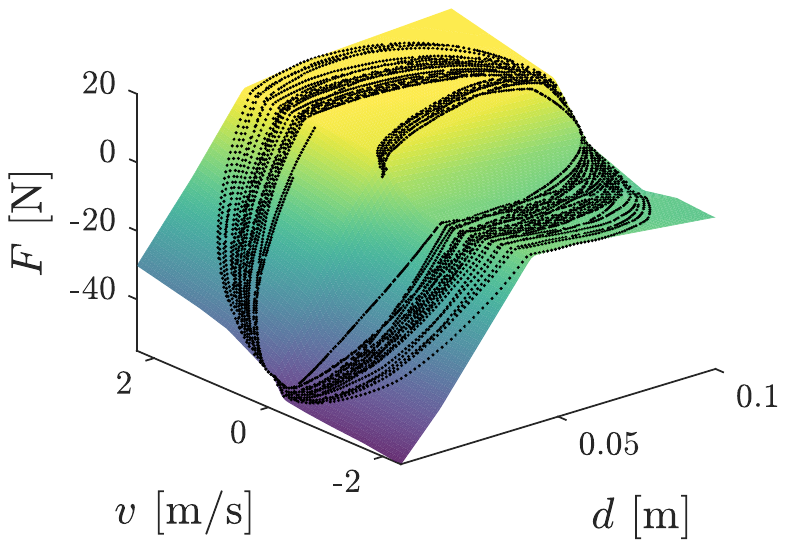}
		\label{fig:NN_dv}
	\end{subfigure}
	\vspace{-5pt}
	\caption{Extracted ReLU network $\eta$ for different model parameters $\vectorstyle{\alpha}$. The black dots indicate the data that were passed through the network during training. }
	\label{fig:NNsurf}
\end{figure}

The identification of the physical parameters in $\vectorstyle{p}=\{m_3,J_1,B_m\}$ are indicative for the convergence of $\eta$ because both parameters $\vectorstyle{p}$ and $\vectorstyle{\alpha}$ are simultaneously optimized. Therefore, we analyze the information captured by $\eta$ after convergence of the NNAP model $\mathcal{M}$ to determine if the unknown load interactions $\mathcal{P}$ can be discoverd. At initialization, the neural network has no physical meaning (no resemblance with the actual $\mathcal{P}$) because the neural network is defined by random learnable parameter values $\vectorstyle{\alpha}_0$. The extracted neural network having as inputs $d$ and $v$ and as output $z=F$ is depicted in Fig. \ref{fig:NNini_dv}. In that phase of training, completely wrong estimations of $z$ are used as the input to ($\ref{eq: phys_general}$), resulting in high initial loss. After convergence to a predictive NNAP model $\mathcal{M}$ (or $\mathcal{R}$), the ReLU network $\eta$ with optimized parameter values $\vectorstyle{\alpha}^\ast$ can be extracted, see Fig. \ref{fig:NN_dv}.  

\begin{figure}[h]
	\centering
	\begin{subfigure}[H]{.49\columnwidth}
		\caption{$\eta(\vectorstyle{q};\vectorstyle{\alpha}^*)$}
		\includegraphics[trim=6.5cm 11.9cm 6.6cm 12.1cm,clip=true,width=1\columnwidth]{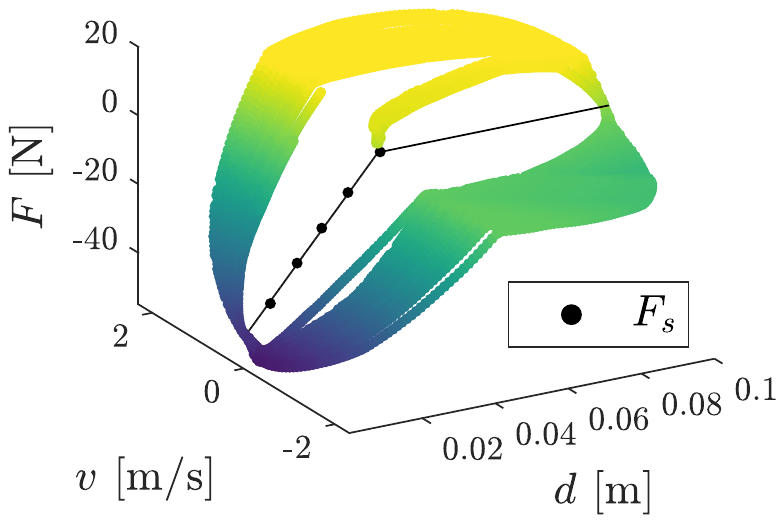}
		\label{fig:NN_scatter_dv}
	\end{subfigure}
	\begin{subfigure}[H]{.49\columnwidth}
		\caption{$\eta(\vectorstyle{q};\vectorstyle{\alpha}^*)$}
		\includegraphics[trim=6.5cm 11.9cm 7cm 12.1cm,clip=true,width=1\columnwidth]{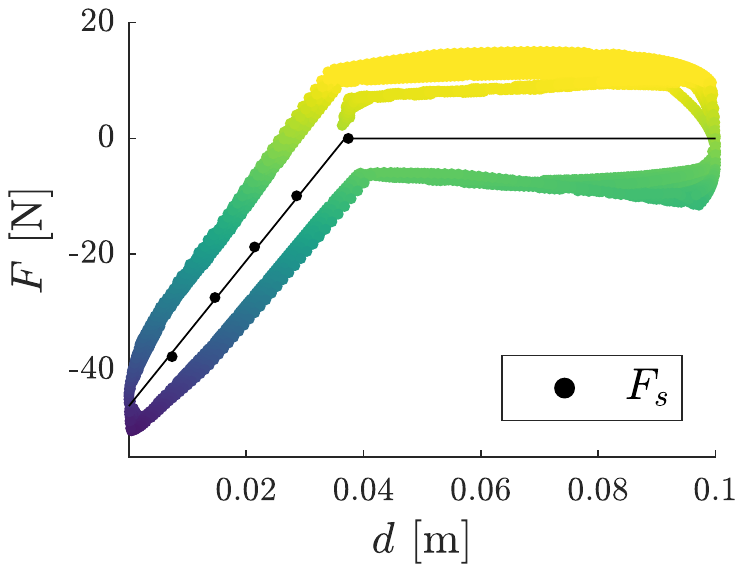}
		\label{fig:NN_scatter_d}
	\end{subfigure}
	\vspace{-5pt}
	\caption{Extracted ReLU network $\eta$ evaluated on the training dataset.}
	\label{fig:physInsight}
\end{figure}

The optimized ReLU network $\eta$ is evaluated on the training dataset, resulting in Fig. \ref{fig:NN_scatter_dv}. The model $\eta$ is only precise locally around the area because data-driven models are inherently difficult in extrapolation capabilities. In \cite{DeGroote2019}, we show for numerical simulation data that the ReLU network $\eta$ indeed converges towards the real force relation $F$ at these explored areas. The corresponding side view, depicted in Fig. \ref{fig:NN_scatter_d}, illustrates the similarity between the empirically validated compression spring force $F_s$ (see also Fig. \ref{fig:springLaw}) and the identified ReLU network $\eta$. From Fig. \ref{fig:NN_scatter_d}, we can deduce a spring law together with friction phenomena. 

\begin{figure}[h]
	\centering
	\begin{subfigure}[H]{.49\columnwidth}
		\caption{spring: $\eta_c$}
		\includegraphics[trim=6.5cm 11.9cm 6.5cm 12.1cm,clip=true,width=1\columnwidth]{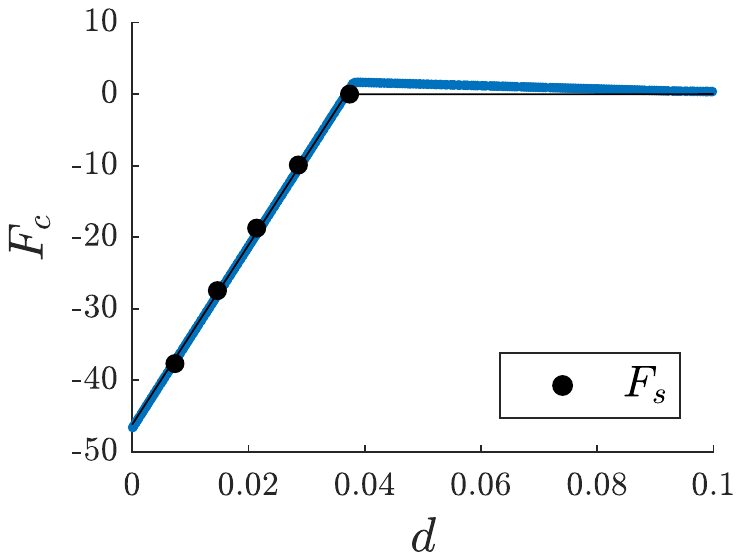}
		\label{fig:Fcon_SpringLoad}
	\end{subfigure}
	\begin{subfigure}[H]{.49\columnwidth}
		\caption{without spring: $\eta_c$}
		\includegraphics[trim=6.5cm 11.9cm 6.5cm 12.1cm,clip=true,width=1\columnwidth]{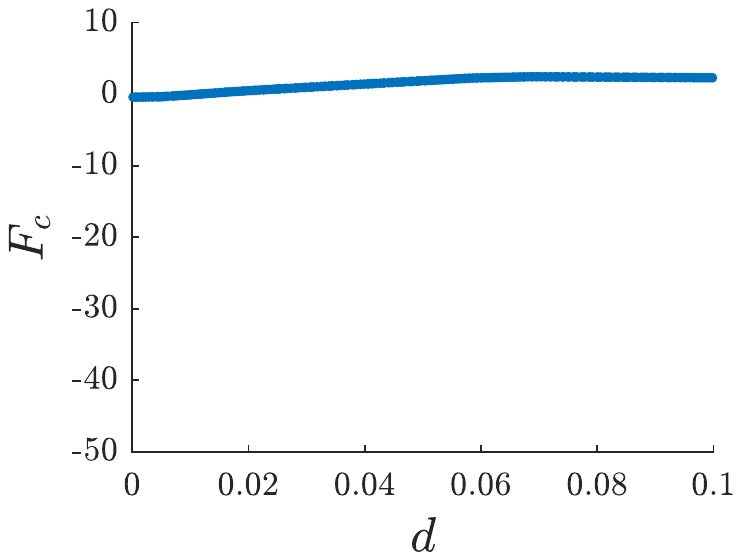}
		\label{fig:Fcon_noLoad}
	\end{subfigure}
	\begin{subfigure}[H]{.49\columnwidth}
		\caption{spring: $\eta_{nc}$}
		\includegraphics[trim=6.5cm 11.9cm 6.5cm 12.1cm,clip=true,width=1\columnwidth]{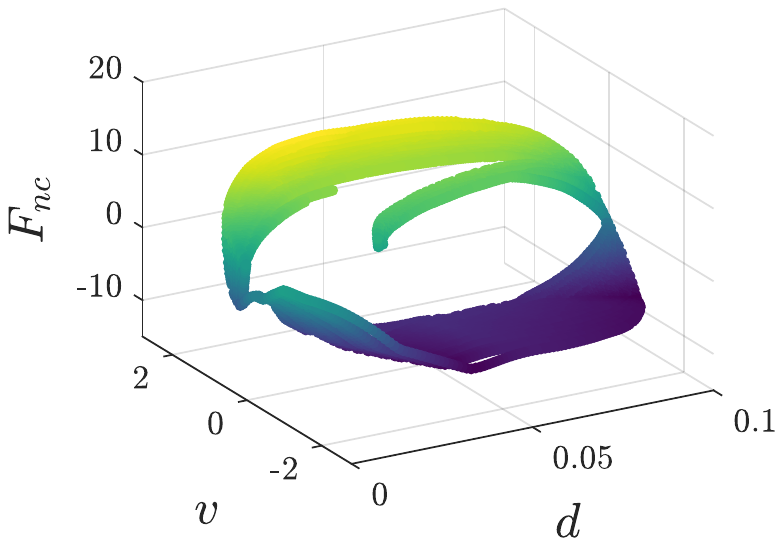}
		\label{fig:Fnoncon_SpringLoad}
	\end{subfigure}
	\begin{subfigure}[H]{.49\columnwidth}
		\caption{without spring: $\eta_{nc}$}
		\includegraphics[trim=6.5cm 11.9cm 6.5cm 12.1cm,clip=true,width=1\columnwidth]{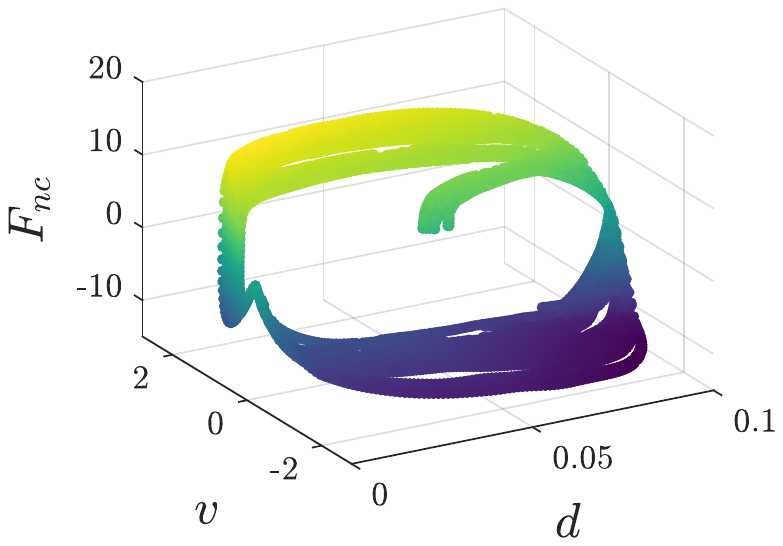}
		\label{fig:Fnoncon_noLoad}
	\end{subfigure}
	\begin{subfigure}[H]{.49\columnwidth}
		\caption{spring: $\eta_{nc}$}
		\includegraphics[trim=6.5cm 11.9cm 6.5cm 11.9cm,clip=true,width=1\columnwidth]{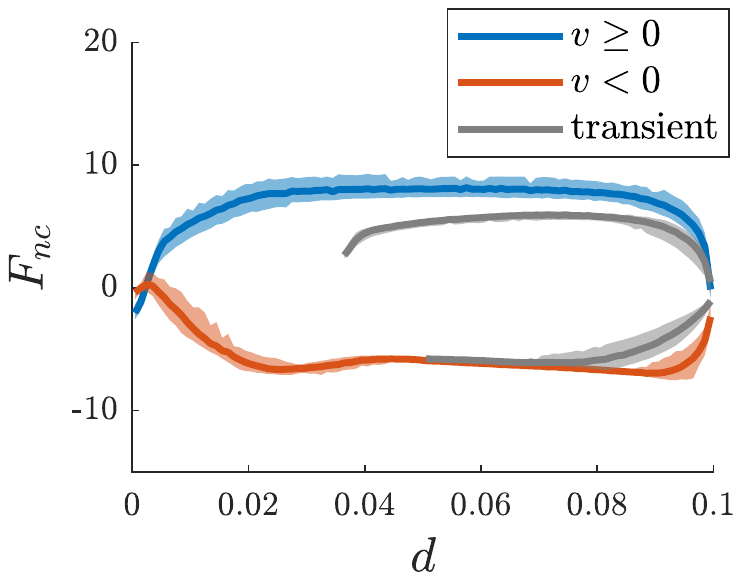}
		\label{fig:Fnoncond_SpringLoad}
	\end{subfigure}
	\begin{subfigure}[H]{.49\columnwidth}
		\caption{without spring: $\eta_{nc}$}
		\includegraphics[trim=6.5cm 11.9cm 6.5cm 11.9cm,clip=true,width=1\columnwidth]{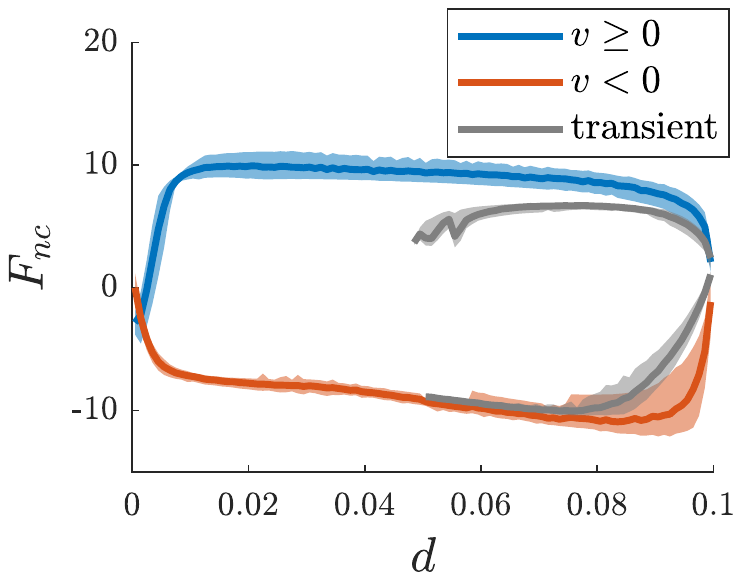}
		\label{fig:Fnoncond_noLoad}
	\end{subfigure}
	\vspace{-15pt}
	\caption{Extracted ReLU networks $\eta_c$ and $\eta_{nc}$ evaluated on the  training dataset, capturing respectively conservative and nonconservative forces.}
	\label{fig:FconAndNonCon}
	\vspace{-20pt}
\end{figure}

The discrepancy between the discovered force relation $\eta$ and the spring force $F_s$ raises the interest to elaborate further on the interpretability of $\eta$. The overall horizontal force acting on the slider is, therefore, considered as a summation $F=F_c +F_{nc}$ of a conservative ($F_c$) and non-conservative ($F_{nc}$) force.  A conservative force is characterized by the property that the work done by moving the object between two points is independent of the taken path.  In practice, we consider a conservative force $F_c(d)$, being only dependent on the position $d$ of the slider. Furthermore, the dissipative force $F_{nc}(d,v)$ relies on both the position $d$ and linear speed $v$. Therefore, we replaced the ReLU network $z=\eta(d,v)$ with a new relation $z=\eta_c(d) + \eta_{nc}(d,v)$ deploying a summation of two separate feedforward ReLU networks $\eta_c(d)$ and $\eta_{nc}(d,v)$ that are trained in parallel within the NNAP model $\mathcal{M}$. The solution of $\eta_c(d)$ and $\eta_{nc}(d,v)$ is, however, not unique because $\eta_{nc}(d,v)$ includes the inputs of $\eta_{c}(d)$. In practice, an additional $L_2$ regularization term ($c\cdot \eta_{nc}^2$), with regularization parameter $c$, was added to the loss function. Fig. \ref{fig:FconAndNonCon} illustrates the identified network components for an experimentally tuned regularization parameter $c=10^{-6}$. The obtained results are the average predictions of $\eta_{c}$ and $\eta_{nc}$ evaluated by the training data for 10 converged NNAP models $\mathcal{M}$. The ReLU network $\eta_c$ is able to accurately predict the behavior of the conservative spring force $F_s$, as can be observed in Fig. \ref{fig:Fcon_SpringLoad}. The regularization was able to successfully address the spring force to the ReLU network $\eta_c$. The remaining unknown dynamics attributed to the $\eta_{nc}$ are shown in Fig. \ref{fig:Fnoncon_SpringLoad}. From the corresponding side view in Fig. \ref{fig:Fnoncond_SpringLoad}, we are able to discover the friction pattern that is dependent on the direction of the slider. 

To validate the methodology that was able to retrieve physical insights, the aforementioned experiment was repeated on the experimental setup having no compression spring. Hence, the state-dependent conservative force is removed. From Fig. \ref{fig:Fcon_noLoad}, it can be seen that the discovered conservative force model $\eta_c$ trends towards zero. The non-conservative force can be extracted from $\eta_{nc}$ and is depicted in Figs. \ref{fig:Fnoncon_noLoad} and \ref{fig:Fnoncond_noLoad}. The non-conservative forces exhibit similar behavior as those shown in \ref{fig:Fnoncon_SpringLoad} and \ref{fig:Fnoncond_SpringLoad}. These results suggest the ability of NNAP models to provide insights into the on-going conservative and non-conservative forces of a mechatronic system interacting with its environment.  

\section{Conclusion}  \label{sec:Conclusion}
This paper presents neural network augmented physics (NNAP) models that can learn the behavior of systems for which the dynamics are only partially known. Data-driven modeling techniques, here neural networks, are inserted in the model to compensate for the unmodeled interactions, without requiring direct measurements of these unknown phenomena. The method is validated on a slider-crank mechanism that exhibited unknown load interactions, induced by a spring and additional friction phenomena. The NNAP model is identified by simultaneously optimizing the neural network and the physical parameters, solely by using state and control input measurements. Therefore, automatic differentiation is used to calculate the gradient information through both the custom physics-inspired and neural layers. The obtained results of the NNAP model are twofold. First, the modeling formalism enabled robust and accurate multistep predictions of a system that was a priori only partially known. Furthermore, both a feedforward and recurrent implementation of the modeling formalism are benchmarked, for which the latter was able to reduce the prediction error by up to 33.4\% on test data. Secondly, the methodology provided a means to discover new insights of the unknown system properties. Results have shown that the NNAP methodology applied to a slider-crank mechanism can uniquely identify uncorrelated sets of physical parameters, while optimizing the neural network. In turn, the neural network, which was extracted from the NNAP model after convergence, accurately captured the state dependent load interactions. The presented methodology, however, heavily relies on the availability of accurate physical models describing the known dynamics. Furthermore, a good understanding of the influence of the unknown phenomena within the overall model is essential to assure physically interpretable results. It should be noted that for more complex cases, the flexibility of the neural network can take over the identification process, resulting in non-unique solutions. Further research is, therefore, required for discovering unknown phenomena using more elaborate regularization strategies.

\appendices

\section{Physics-Based Model} \label{App:Model}
\begin{figure}[h]
	\centering
	\includegraphics[trim=4.7cm 4.1cm 4.6cm 7.5cm,clip=true,angle=-90,width=0.9\columnwidth]{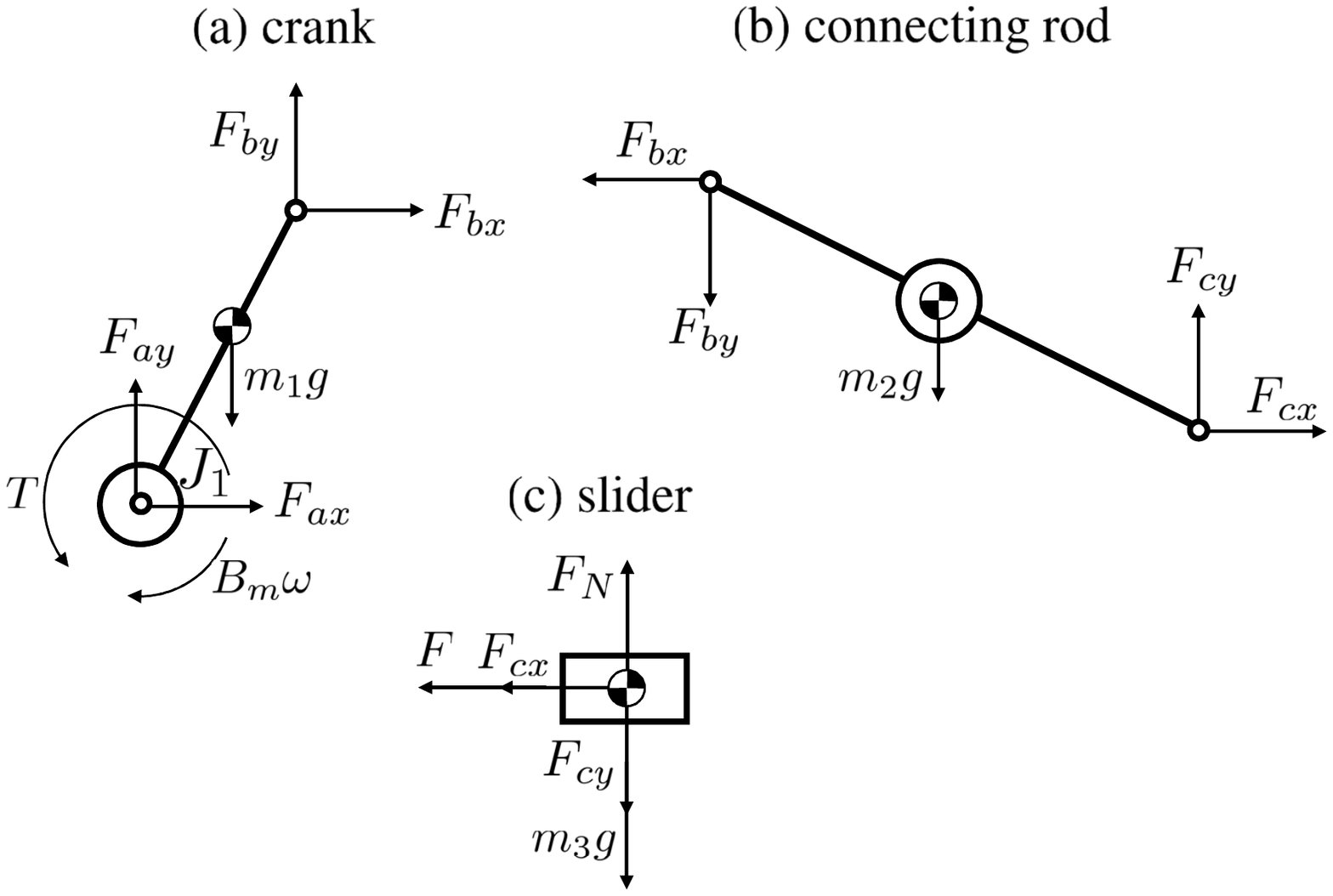}
	\vspace{-10pt}
	\caption{Multibody scheme of slider-crank mechanism.}
	\label{fig:multibodyDyn}
	\vspace{-5pt}
\end{figure}

The physics-based dynamical model of the slider-crank mechanism is derived by considering the  simplified multibody system in Fig. \ref{fig:multibodyDyn}. The variable $J_1$ includes the rotational inertia of both motor and crank with respect to the rotation axis. The rotational inertia $J_2$ of the connection rod is approximated by $J_2=\frac{1}{3}l_2m_2^2$ in its center of mass (CM). Subsequently, this implies that we take the $r_2=\frac{l_2}{2}$ in Fig. \ref{fig:geometrical}. By analogy, we approximate the CM of the crank by $r_1=\frac{l_1}{2}$. Furthermore, we define $r'=l-r$ and let the subscripts $r_x$ and $r_y$ indicate the projection of $r$ on, respectively, the $x$-axis and $y$-axis.
\begin{equation}
\label{eq:Phys1}
\begin{aligned}
m_1 \ddot{x}_1 & = F_{ax}+ F_{bx} \\
m_1 \ddot{y}_1 & = -m_1g + F_{ay} + F_{by} \\
J_1 \dot{\omega} & = T -B_m \omega -r_{1x}m_1g+l_{1x}F_{by} - l_{1y}F_{bx} \\ 
m_2 \ddot{x}_2  & =-F_{bx} +  F_{cx} \\
m_2 \ddot{y}_2  & =-F_{by} + F_{cy} -m_2g \\
J_2 \ddot{\phi} & = -r'_{2x}F_{by}-r'_{2y}F_{bx}-r_{2x}F_{cy}-r_{2y}F_{cx} \\ 
0 & = -F_{cy}-m_3g+F_N\\
m_3 \dot{v} & = -F_{cx}-F \\
\end{aligned}
\end{equation}
The aforementioned equations are combined with the geometrical relations, illustrated in Fig. \ref{fig:geometrical}, towards a physics-inspired model $\dot{\vectorstyle{x}}=\vectorstyle{f}(\vectorstyle{x},u,F)$ with state $\vectorstyle{x}=\begin{bmatrix} \theta & \omega \end{bmatrix}^T$ and control input $u=T$. The highly nonlinear expression of $\vectorstyle{f}$ is derived via symbolic solvers and is due to its complexity not mentioned in this paper. 

\begin{table}[h]
	\centering
	\caption{Identified parameters of slider-crank setup.}
	\begin{tabular}{|c|c|c c|}
		\hline
		$m_1$ & mass of crank &  0.23 & [\si{kg}] \\
		$m_2$ & mass of connection rod &  0.35 & [\si{kg}] \\
		$m_3$ & mass of slider & 0.77 & [\si{kg}] \\		
		$l_1$ & length of crank & 0.05 & [\si{m}] \\
		$l_2$ & length of connection rod & 0.29 & [\si{m}] \\
		$J_1$ & inertia motor + crank  & 0.0034 &[\si{kg.m^2}]\\ 
		$B_m$ & friction coefficient of motor&	0.007   & [$\frac{\si{N.m}}{\si {rad/s}}$] \\	
		\hline
	\end{tabular}
	\label{table:parReal}
	\vspace{-5pt}
\end{table}

\section{Sensitivity Analysis} \label{App:SensitivityAnalysis}
\begin{figure}[h]
	\centering
	\vspace{-10pt}
	\includegraphics[trim=4.3cm 12.5cm 5cm 13cm,clip=true,width=0.8\columnwidth]{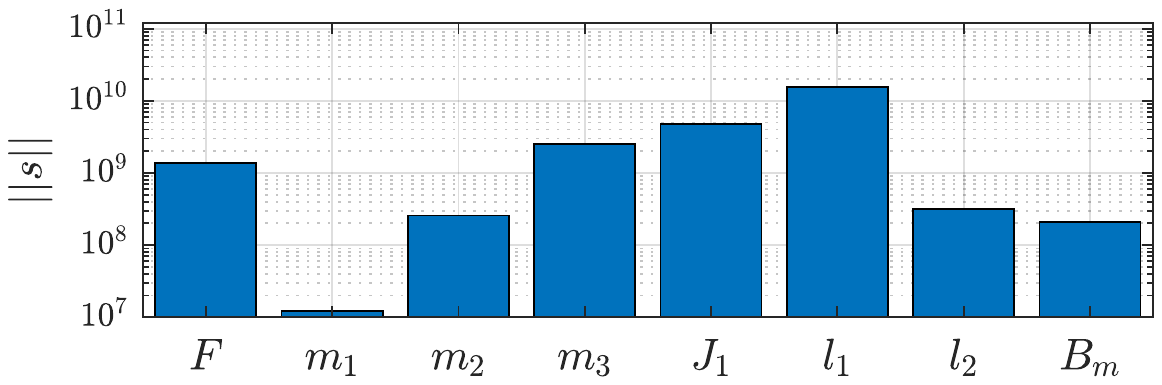}
	\vspace{-5pt}
	\caption{Local sensitivity analysis of unknown force interaction $F$ and physical parameters $\vectorstyle{p}$.}
	\vspace{-10pt}
	\label{fig:Fish_diag}
\end{figure}

\begin{figure}[H]
	\centering
	\vspace{-10pt}
	\includegraphics[trim=4.5cm 12cm 5cm 12.5cm,clip=true,width=0.8\columnwidth]{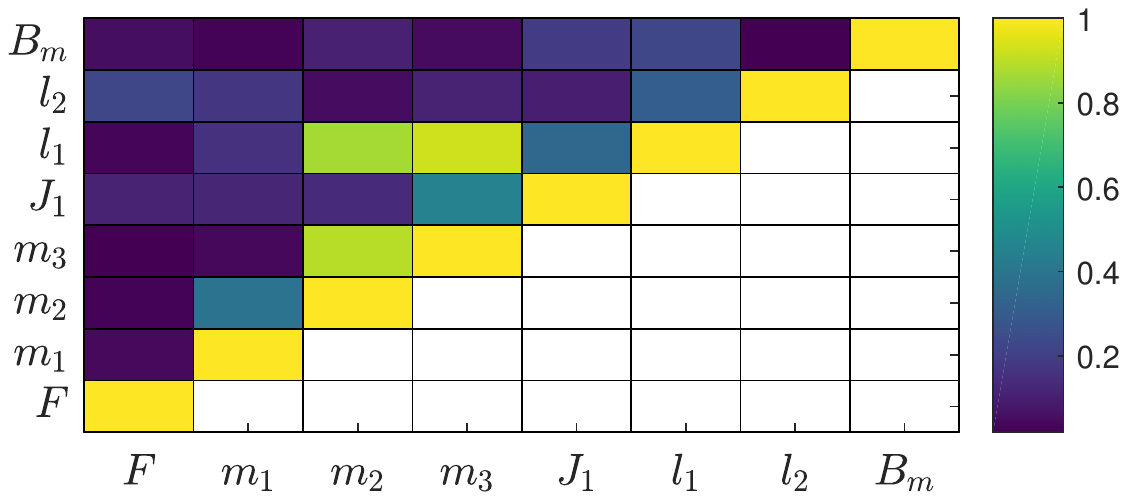}
	\vspace{-10pt}
	\caption{Correlation matrix $\matrixstyle{Q}$ of the sensitivity vectors $\vectorstyle{s}^N_j$.}
	\vspace{-5pt}
	\label{fig:Fish_nondiag}
\end{figure}

\begin{figure}[H]
	\centering
	\vspace{-5pt}
	\includegraphics[trim=5.2cm 12.8cm 5cm 13.5cm,clip=true,width=1\columnwidth]{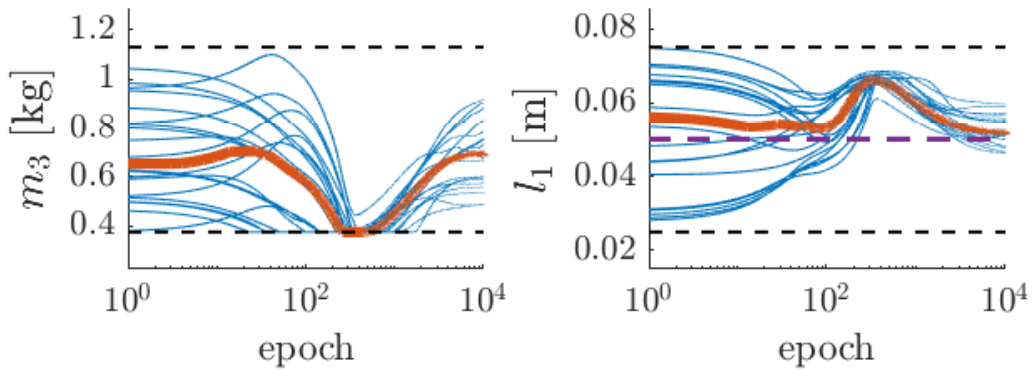}
	\vspace{-15pt}
	\caption{Convergence history of $l_1$ and $m_3$ when including both in the parameter set $\vectorstyle{p}=\{l_1,m_3,J_1,B_m\}$, simultaneously optimized with the neural network parameters $\vectorstyle{\alpha}$.}
	\vspace{-10pt}
	\label{fig:parOptim_l1andm3}
\end{figure}

The identified physical parameters $\vectorstyle{p}$ of the slider-crank mechanism
are depicted in Table \ref{table:parReal}. The parameters that could not be determined by optimizing the NNAP model are measured experimentally or determined via CAD drawings. Because some parameter combinations may lead to the same model output, non-unique solutions may, however, appear. The highly nonlinear nature of the proposed slider-crank model makes a structural identifiability analysis non trivial. However, based on \cite{ioslovich2010identification}, we implemented approximation techniques indicative for local identifiability around the optimal solution. We elaborate on the derivative function $f$ evaluated for a sample $k$ as $\dot{\omega}_k = f(\theta_k, \omega_k, T_k, \hat{F}_k ; \vectorstyle{p})$. In practice, all measurement trajectories are aggregated into one dataset of $K$ samples. The converged neural network $\eta$ is deployed on each measurement sample $k \in  \{1 , \ldots, K \}$ to generate the corresponding estimations $\hat{F}_k$ of the unknown force interaction. Next, we derive the sensitivity matrix $\matrixstyle{S}$ in which we quantify the impact on the model output for modifications on $F$ and physical parameters $\vectorstyle{p} \in \mathbb{R}^{r}$.
\begin{equation*}
\matrixstyle{S}=\begin{bmatrix}  \vectorstyle{s}_1   \cdots  \vectorstyle{s}_{r+1}   \end{bmatrix}
=
\begin{bmatrix}
\left. {\frac{\partial f}{\partial F}} \right|_{k=1}%
& 
\left. {\frac{\partial f}{\partial p_1}} \right|_{k=1}%
& \ldots &
\left. {\frac{\partial f}{\partial p_r}} \right|_{k=1}\\
\vdots & \vdots & \ddots& \vdots \\
\left. {\frac{\partial f}{\partial F}} \right|_{k=K}%
& 
\left. {\frac{\partial f}{\partial p_1}} \right|_{k=K}%
& \ldots &
\left. {\frac{\partial f}{\partial p_r}} \right|_{k=K}\\
\end{bmatrix}
\end{equation*}
Again, we consider normalized parameters in order to develop a fair scaling to cope with the different sizes of magnitude. The  $L_2$  norm of each sensitivity vector $\vectorstyle{s}_j $ measures the influence of the corresponding feature $j$ to the model output. Figure \ref{fig:Fish_diag} illustrates the high sensitivity of the parameters  $l_1$, $J_1$ and $m_3$, justifying the choice to include them in the optimization set $\vectorstyle{p}$. In addition, the parameter $B_m$ is added because this variable is unknown and cannot be measured. The high sensitivity of $F$, which is representative for the output of the neural network $\eta$, is related to the convergence of $\eta$. 
Although a set of variables can be highly sensitive, if parameter combinations have the same influence on the model output, they cannot be uniquely determined. Therefore, we construct the normalized sensitivity matrix $\matrixstyle{S}_N$ that contains the normalized column vectors $\vectorstyle{s}_j^N=\vectorstyle{s}_j/||\vectorstyle{s}_j||$ for $j \in \{1,..,r+1\}$.  The corresponding correlation matrix $\matrixstyle{Q}=\matrixstyle{S}_N^T \matrixstyle{S}_N$ exhibits diagonal elements equal to one and results in the graph shown in Fig. \ref{fig:Fish_nondiag}. This result demonstrates that the physical parameters are barely correlated with the output of the neural network represented by $F$. However, we can observe, for instance, a high correlation of parameters $l_1$ with $m_3$ (and $m_2$) on the model output. When considering parameter set $\vectorstyle{p}=\{l_1,m_3,J_1,B_m\}$, we can observe from Fig. \ref{fig:parOptim_l1andm3} that they indeed do not converge well in the optimization process.

\bibliography{references} 
\bibliographystyle{IEEEtran}
\vspace{-20pt}
\begin{IEEEbiography}[{\includegraphics[width=0.8in,height=1.25in,clip,keepaspectratio]{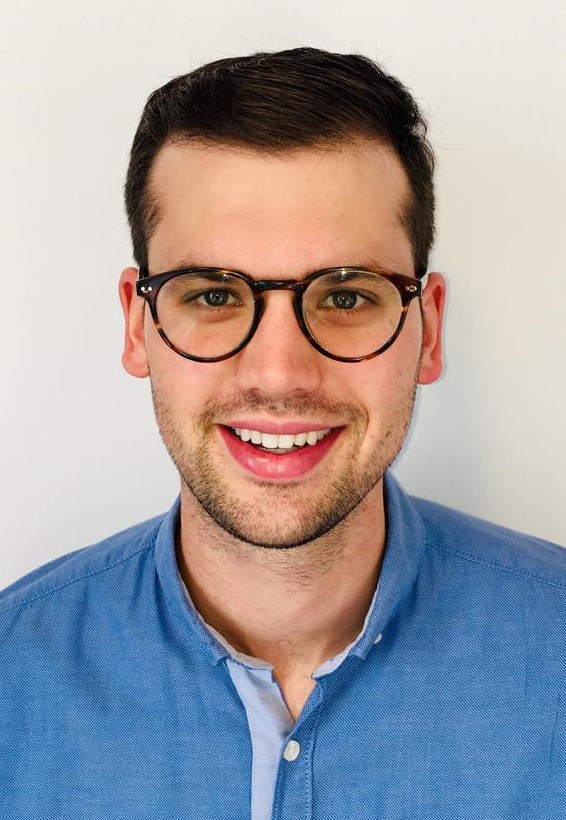}}]{Wannes De Groote} Wannes De Groote received a M.Sc. degree in electromechanical and control engineering from Ghent University, Ghent, Belgium, in 2017. In August 2017, he joined the Department of Electromechanical, Systems and Metal Engineering, Ghent University, as a Ph.D. Student. He received a personal grant strategic basic research (SB) from Research Foundation - Flanders (FWO) in 2019. His research interests mainly concern simulation and modeling of dynamic systems by combining physical and data-driven approaches.
\end{IEEEbiography}
\vspace{-20pt}
\begin{IEEEbiography}[{\includegraphics[width=0.8in,height=1.25in,clip,keepaspectratio]{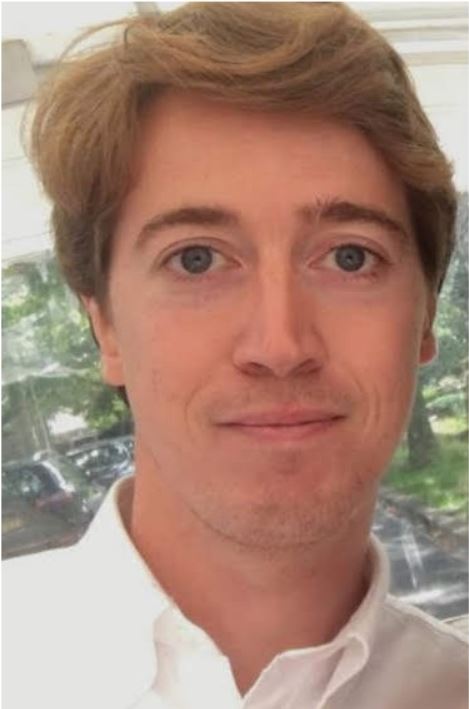}}]{Edward Kikken} 
Edward Kikken obtained his master's degree in mechanical engineering - focused on control theory - at the Technical University of Eindhoven in 2013. His professional career began as an R\&D engineer focused on advanced motion control for offshore application at Allseas. Since 2017, he has been working for the Flemish strategic research institute Flanders Make, currently as a senior researcher in systems and control, signal processing, optimization, and machine learning.
\end{IEEEbiography}
\vspace{-20pt}
\begin{IEEEbiography}[{\includegraphics[width=0.8in,height=1in,clip,keepaspectratio]{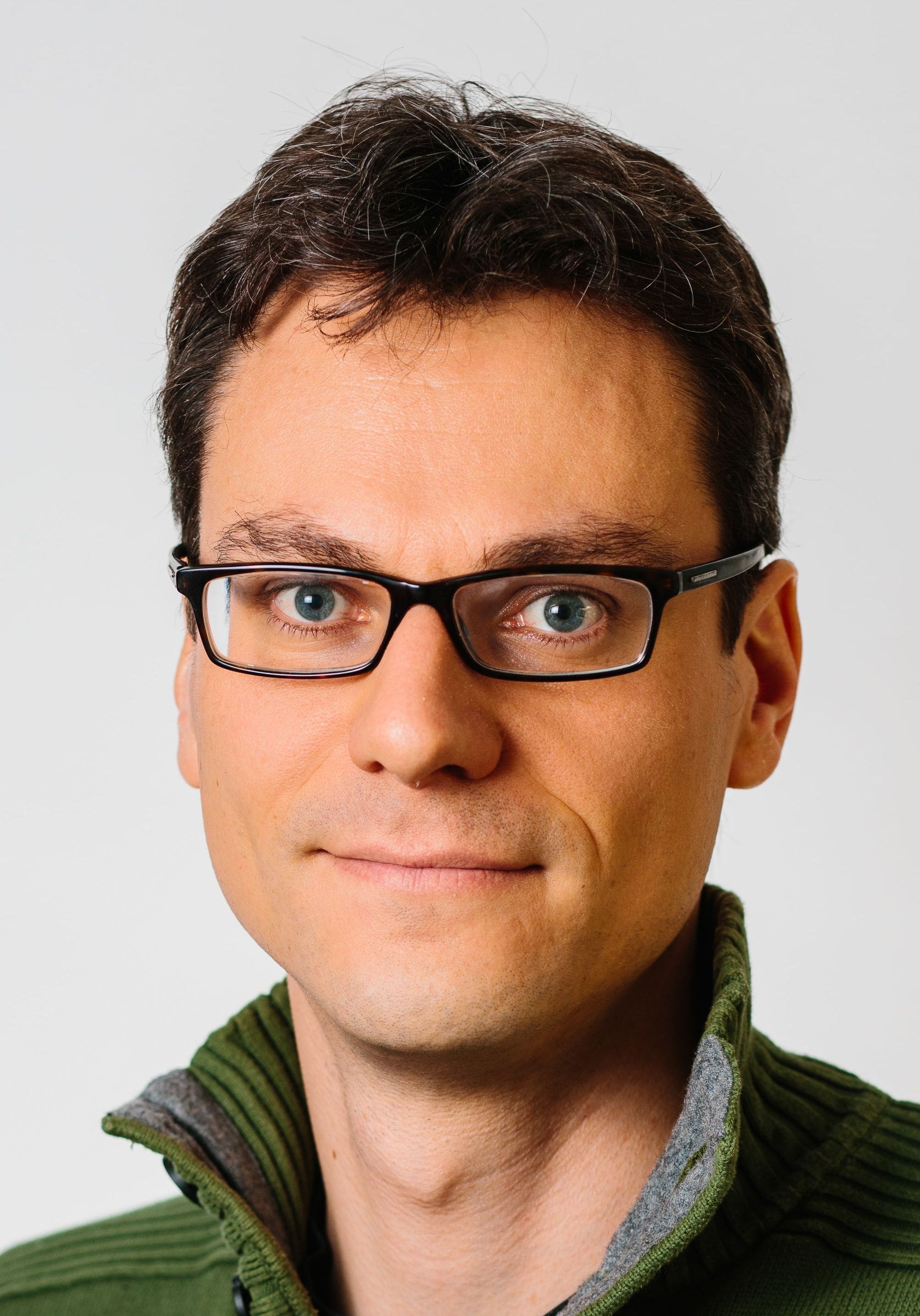}}]{Erik Hostens} 
Erik Hostens earned a master's degree magna cum laude in electromechanical engineering (datamining and automation) in 2003, and a PhD in quantum information theory in 2007, both at KULeuven. He worked for a few years in R\&D on advanced process control for Siemens. Since 2010, he has been employed at Flanders Make, currently as a senior researcher in systems and control, estimation, optimization, and machine learning.
\end{IEEEbiography}
\vspace{-20pt}
\begin{IEEEbiography}[{\includegraphics[width=0.8in,height=1.25in,clip,keepaspectratio]{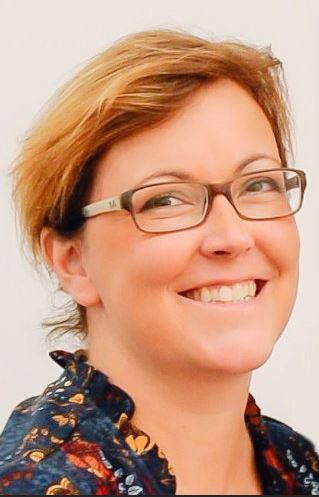}}]{Sofie Van Hoecke}
	Sofie Van Hoecke is an Associate Professor at IDLab, Ghent University-imec. She and her team are active on the study and development of innovative hybrid machine learning solutions combining machine learning with semantics, expert knowledge and/or physical knowledge, efficient and accurate characterization, modeling and training, and testing and optimization, with applications in predictive maintenance and predictive healthcare. 
\end{IEEEbiography}
\vspace{-20pt}
\begin{IEEEbiography}[{\includegraphics[width=0.8in,height=1.25in,clip,keepaspectratio]{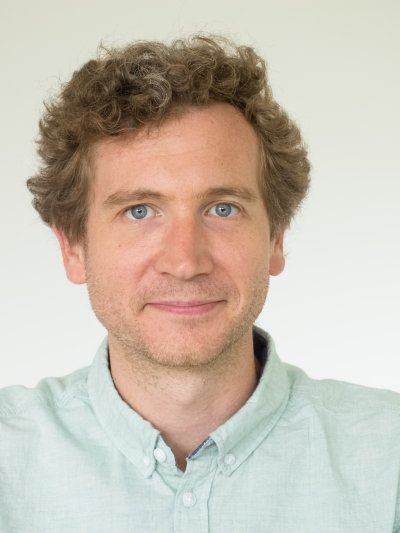}}]{Guillaume Crevecoeur} 
	Guillaume Crevecoeur (°1981) received a MSc and a PhD degree in Engineering Physics from Ghent University in 2004 and 2009, respectively. In 2009, he became a postdoctoral fellow of the Research Foundation Flanders. In 2014, he was appointed Associate Professor at the Faculty of Engineering and Architecture of Ghent University. His research interests are the modeling, optimization, and control of dynamical mechatronic and industrial robotic systems.	
\end{IEEEbiography}
\newpage
\end{document}